\documentclass[12pt,preprint]{aastex}

\newcommand{\myemail}{krimm@milkyway.gsfc.nasa.gov}

\slugcomment{Submitted to ApJ}

\shorttitle{GRB 060714}
\shortauthors{Krimm et al.}

\begin{document}


%\date{\today}
\title{GRB 060714: No Clear Dividing Line Between Prompt Emission and X-ray Flares}


\author{H. A. Krimm\altaffilmark{1,2}, J. Granot\altaffilmark{3}, 
 F. E. Marshall\altaffilmark{4}, M. Perri\altaffilmark{5}, S. D. Barthelmy\altaffilmark{4}, 
D. N. Burrows\altaffilmark{6}, N. Gehrels\altaffilmark{4}, 
 P. M\'{e}sz\'{a}ros\altaffilmark{6},
D. Morris\altaffilmark{6}}
%\affil{Astronomy Department, University of California,
 %   Berkeley, CA 94720}

%\author{C. D. Biemesderfer\altaffilmark{4,5}}
%\affil{National Optical Astronomy Observatories, Tucson, AZ 85719}
\email{\myemail}

%\and

%\author{R. J. Hanisch\altaffilmark{5}}
%\affil{Space Telescope Science Institute, Baltimore, MD 21218}


\altaffiltext{1}{CRESST and NASA Goddard Space Flight Center, Greenbelt,
  MD 20771, USA}
\altaffiltext{2}{Universities Space Research Association, 10211
Wincopin Circle, Suite 500, Columbia, MD 21044, USA}
\altaffiltext{3}{KIPAC, Stanford University, P.O. Box 20450, MS 29,
Stanford, CA 94309, USA}
\altaffiltext{4}{NASA Goddard Space Flight Center, Greenbelt,
  MD 20771, USA}
\altaffiltext{5}{ASI Science Data Center, Via Galileo Galilei, I-00044
Frascati, Italy}
\altaffiltext{6}{Department of Astronomy and Astrophysics, 525 Davey
Lab., Pennsylvania State University, University Park, PA 16802, USA}

\begin{abstract}
%\noindent  
The long gamma-ray burst GRB~060714 was observed to exhibit a series of five X-ray flares beginning $\sim 70\;$s after the burst trigger $T_0$\ and continuing until $\sim T_0+200\;$s. The first two flares were detected by the Burst Alert Telescope (BAT) on the {\em Swift} satellite, before {\em Swift} had slewed to the burst location, while the last three flares were strongly detected by the X-Ray Telescope (XRT) but only weakly detected by the BAT. This burst provides an unusual opportunity to track a complete sequence of flares over a wide energy range. The flares were very similar in their light curve morphology, showing power-law rise and fall components, and in most cases significant sub-structure. The flares also showed strong evolution with time, both spectrally and temporally. The small time scale and large amplitude variability observed are incompatible with an external shock origin for the flares, and support instead late time sporadic activity either of the central source or of localized dissipation events within the outflow. We show that the flares in GRB 060714 cannot be the result of internal shocks in which the contrast in the Lorentz factor of the colliding shells is very small, and that this mechanism faces serious difficulties in most Swift GRBs. The morphological similarity of the flares and the prompt emission and the gradual and continual evolution of the flares with time makes it difficult and arbitrary to draw a dividing line between the prompt emission and the flares.

\end{abstract}





\section{Introduction}

\noindent One of the most surprising findings of the studies of gamma-ray bursts
(GRBs) made with the {\em Swift
Gamma-Ray Burst Explorer} \citep{gehr04} is that nearly half of all bursts show flares, or large
short-lived increases in emission at times after the initial prompt
emission has died away \citep{burro05b,nou06,obr06}.  These flares are
superimposed on either the rapidly decaying tail of the prompt
emission, or the very slowly decaying phase of the early
afterglow, and can involve flux increases of as much as three orders
of magnitude.  Some bursts have single flares, although most bursts
with flares have multiple flaring episodes.  Most flares are at early
times, $t \lesssim 10^3\;$s, although strong flares can occur as late as
$\gtrsim 10^4\;$s after the 
onset of the burst.
	
The origin of GRB flares is still an open question. Evidence is
mounting, however, that the origin of these flares is similar to that
of the prompt GRB emission (i.e. either internal shocks or some other
well localized dissipation process within the ultra-relativistic
outflow), rather than to that of the afterglow emission (i.e. the
external shock going into the ambient medium). This evidence 
includes the multiplicity of flares; their sharp time structure: rapid
rise and decay and sub-peak structure within the flares; the large
increase in flux during the flare; and the hard to soft spectral
evolution of the flares, which is similar to the spectral evolution
found in the prompt emission.

In the context of the internal shocks model \citep{rees94}, flares may
be caused by late-time collisions of shells of relativistic material
that are produced by the central engine with varying Lorentz
factors. These can occur either by late time sporadic activity of the
central source, or by a small relative velocity between shells that
were ejected during the prompt GRB emission \citep[e.g.,][]{burro07}.

Very few GRBs have a
sequence of multiple flares bright enough to be studied in detail.  GRB~060714
\citep{gcn5311} showed a series of five flares starting at $\sim 70$\
s after the start of the burst.  The first three flares were clearly
detected by the {\em Swift} Burst Alert Telescope
\citep[BAT;][]{barth05} and the last three were seen as strong flares by the {\em Swift} X-Ray
Telescope \citep[XRT;][]{hill04, burro05} and as weak flares by the BAT.  
The first two flares occurred while the spacecraft was slewing to or settling at the burst location, so they
were not observable by the XRT.  This burst
provides a rare opportunity to study a rapid sequence of flares across
a large energy range.  The five flares show evidence of hard to soft
spectral evolution as the flares progress and also strong
similarities, in particular sharply resolved temporal features and
large flux increases in each flare.  We show that these results are inconsistent
with an external shock (i.e. afterglow) origin for the flares, and
suggest instead a late-time and lower energy continuation of the
prompt emission, either due to late time intermittent activity of the
central source or via well localized spasmodic dissipation events
within the outflow.

In fact these results suggest that it may no longer be possible to draw a clear distinction 
between what have traditionally been called the prompt emission and the X-ray flares.  
Historically the prompt emission was that detected above $\sim 20$\ keV, and 
generally considered to be due to activity of the central engine.  X-ray flares are
usually detected most strongly at lower energies, so they were considered a completely
separate phenomenon.  Since we show here that (a) the flares of GRB~060714 are very likely
of common origin to the earliest emission from the burst, (b) the flares
are detected above 20~keV, and (c) the flares show a gradual and continual evolution linking them
to the prompt emission, it is quite reasonable to consider the flares a lower energy 
continuation of the same phenomenon as the prompt emission.   However to be consistent with
earlier work we do use the term ``prompt emission" to refer to the emission in the first peak, $\sim -13 - +20$\ s from the burst trigger (see \S~\ref{BAT}), ``flare" to refer to the peaks more than $\sim 70$\ s
after the trigger, and ``afterglow'' to refer to the smooth decay $\gtrsim 300$\ s from the trigger.

In this paper we describe the prompt, flaring and afterglow properties
of GRB~060714, with particular emphasis on the flares.  In
\S~\ref{sec2} we discuss the observations and data analysis in
general, while \S~\ref{flares} focuses on the analysis of the spectral
and temporal properties of the flares. Finally, in \S~\ref{sec4} we
show that the
properties of the flares rule out an external shock (or long lived
reverse shock) origin, and provide some (though not unequivocal) support
for an origin common to that of the prompt emission. In \S 4.3 and Appendix A we exclude internal shocks with a
small contrast in the Lorentz factor as the origin of the flares in GRB~060714, and point out that this mechanism faces serious problems for most
Swift GRBs.

Throughout the paper we have followed the convention $F_{\nu,t} \propto \nu^{-\beta}t^{-\alpha}$,
where the energy spectral index $\beta$\ is related to the photon index $\Gamma\ = \beta\ +\ 1$.  We have 
adopted the standard values of the cosmological parameters:   ${\rm H_0\ =\ 70\ km\ s^{-1}\ Mpc^{-1}}$,
$\Omega_M$\ = 0.27, and $\Omega_{\Lambda}$\ = 0.73.  The phenomenology
of the burst is presented in the observer frame unless otherwise stated. 

\section{Observations and Data Analysis}
\label{sec2}
 
\subsection{Swift-BAT}
\label{BAT}

\noindent At 15:12:00 UT, 14 July 2006, the {\em Swift} Burst Alert
Telescope (BAT) triggered and located on-board GRB~060714 \citep[BAT
trigger=219101;][]{gcn5311}. Unless otherwise specified, times $t$ in
this paper are measured from the BAT trigger time, (UT 15:12:00.3)
hereafter designated T$_0$.  The burst was detected in the part of the
BAT field of view that was 27\% coded, meaning that it was 33.6$\degr$
off-axis and only 27\% of the BAT detectors were illuminated by the
source.  The spacecraft began to slew to the source location at
T$_0+34.9\;$s and was settled at the source location at T$_0+88.1\;$s.

\begin{figure}
\plotone{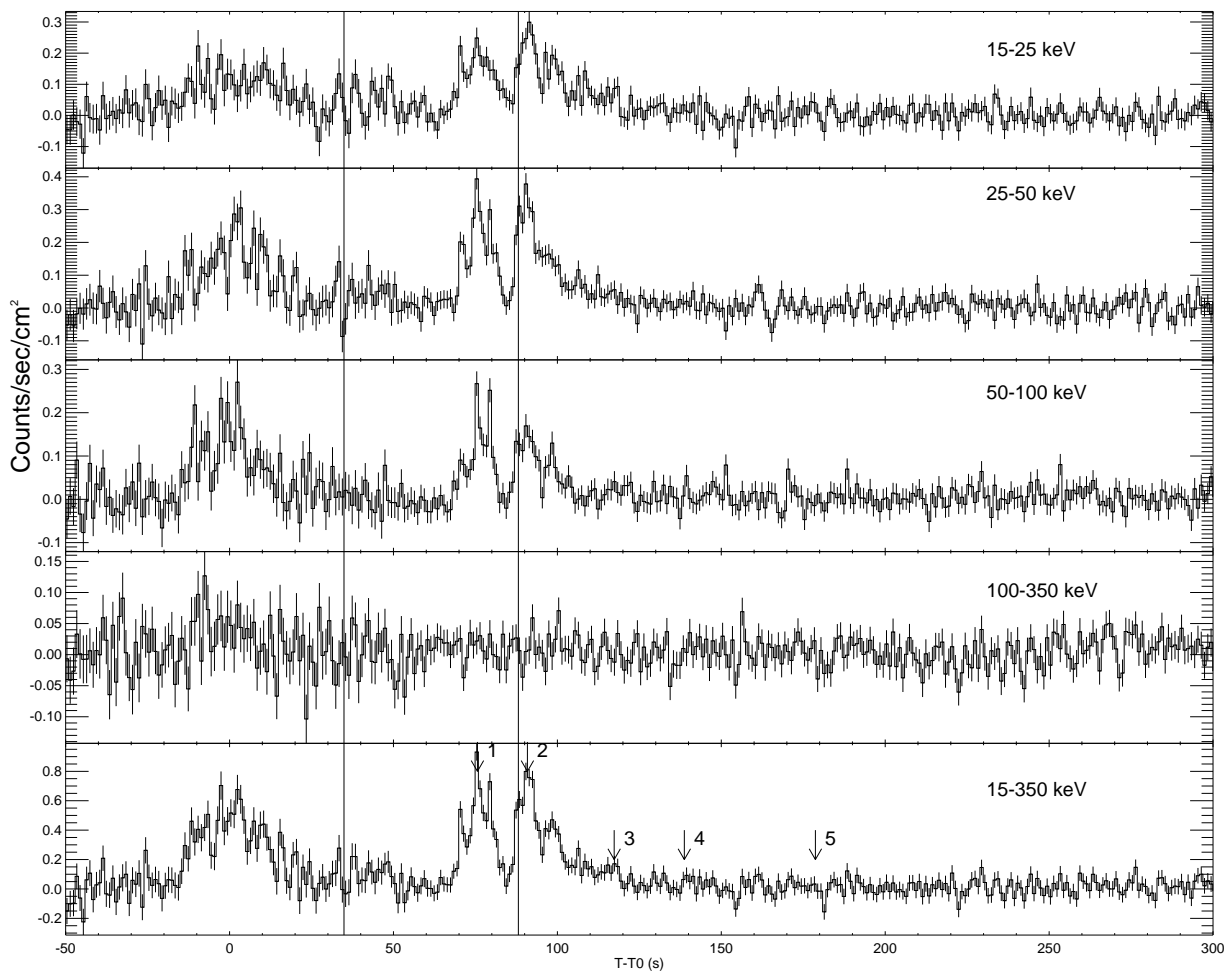}
\caption{The BAT light curve is shown in four energy bands and the sum
of all energy bands.  The vertical bars indicate the start and end of
the spacecraft slew to the burst location.  Note that the count rate
statistical errors are much larger before the slew than after the
slew.  This is because the burst was detected near the edge of the BAT
field of view where only 27\% of the detectors were illuminated.  None
of the apparent sharp structure in the prompt emission is
statistically significant. The arrows in the bottom plot indicate the peak and the
numbering of each of the flares. \label{fig1}}
\end{figure}

\begin{figure}
\plotone{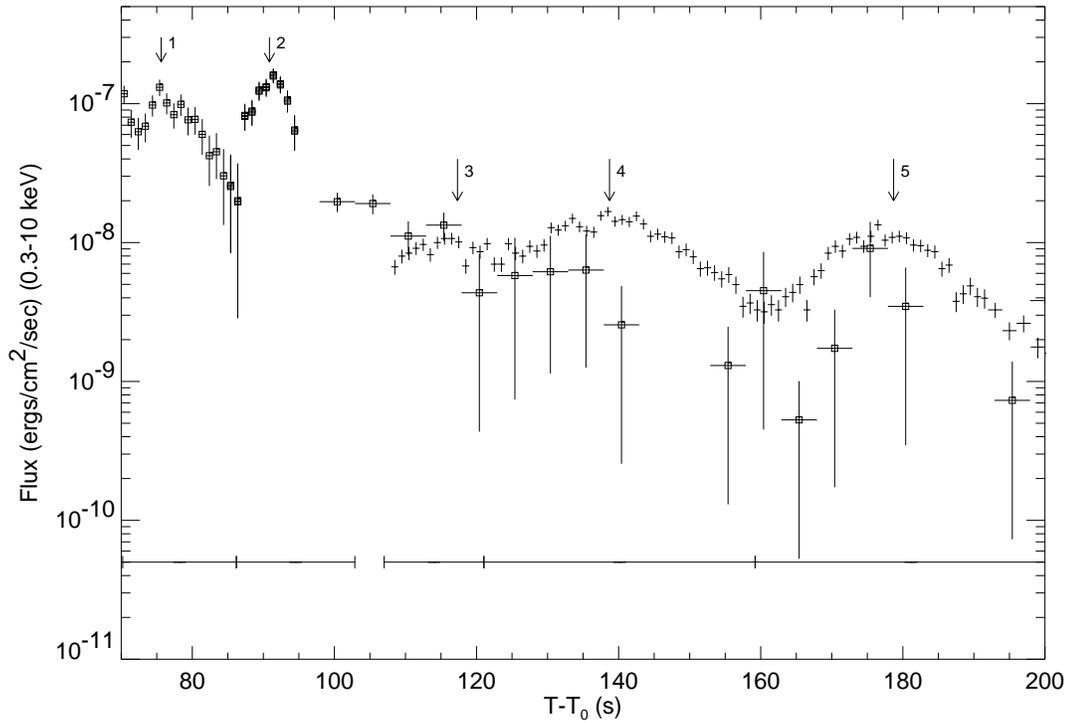}
\caption{The five flares are shown for both BAT (open squares) and XRT
(crosses).  The first XRT flare at T$\;\sim 115\;$s is clearly detected
in the BAT and there appears to BAT emission at the peak of the flare
at T$\;\sim 175\;$s.  The arrows indicate the peak and the
numbering of each of the flares. The time for which the spectral fits to the five
flares were made are indicated by the bars at the bottom of the
plot.\label{fig4}}
\end{figure}

The BAT data for GRB 060714 between T$_0-240\;$s and T$_0+962\;$s were
collected in event mode with 100 $\mu$s time resolution and
$\sim$6~keV energy resolution \citep{gcn5334}.  The data were
processed using standard {\it Swift}-BAT analysis tools and the spectra were
fit using {\sc xspec 11.3}.  Each BAT event was mask-tagged using {\sc
batmaskwtevt} with the best fit source position. Mask-tagging is a
technique in which each event is weighted by a factor representing the
fractional exposure to the source through the BAT coded aperture.  A
weight of +1 corresponds to a fully open detector and a weight of -1
to a fully blocked detector.  Flux from the background and other
sources averages to zero with this method.  All of the BAT GRB light
curves shown in Figures~\ref{fig1}, \ref{fig4}, and \ref{fig5} have
been background subtracted by this method.  This method is effective
even when the spacecraft is moving since complete aspect information
is available during the maneuver.

The mask-weighting is also applied to produce weighted, background
subtracted counts spectra using the tool {\sc batbinevt}.  Since the
response matrix depends on the position of the source in the BAT field
of view, separate matrices are derived for before the slew, after the
slew and for individual segments of the light curve during the slew.

The mask-weighted lightcurves seen in Figure~\ref{fig1} show an
initial triangular-shaped (rising and falling power laws) peak starting at T$_0-15\;$s, peaking at
T$_0$, and ending at T$_0+55\;$s. Given the size of the statistical error
bars in the time before the slew, none of the fluctuations seen in
Figure~\ref{fig1} before the slew are statistically significant.
The initial prompt emission was
followed by two strong flares, the first starting at T$_0+70.4\;$s
with a duration of $\sim 10.5\;$s, and the second starting at
T$_0+87.6\;$s with a duration of $\sim 12\;$s.  Finally there was a
much weaker third flare, starting at T$_0+108.7\;$s with a duration of
$\sim 15.3\;$s.  Taking into account the prompt emission plus the
flares, we derive $T_{90}\ =\ 115 \pm 5\;$s (estimated error including
systematics).  

The time-averaged spectrum from $T_0-13.4\;$s to $T_0+18.0\;$s is best
fit by a simple power-law model. The power law photon index of the
time-averaged spectrum is $\Gamma = -d\log N_{\rm ph}/d\log E_{\rm ph}
= 1.61 \pm 0.13$ ($\chi ^2 = 43.2$ for 59 d.o.f.). The fluence in
the 15-150 keV band is $(1.22 \pm 0.11) \times 10^{-6}$ erg
cm$^{-2}$. The $1\;$s peak photon flux measured from T$_0$+75.42 s
(during the first flare) in the 15-150 keV band is $1.4\pm 0.1$ ph
cm$^{-2}$ s$^{-1}$.   The prompt component does show spectral evolution,
as evidenced by the increasing power law index:   ($T_0-13.4$ to $+2.1\;$s):  $1.47 \pm 0.19$;
($T_0+2.1\;$ to $+18.0\;$s):  $1.61 \pm 0.17$; ($T_0+18.0\;$ to $+70.2\;$s):  $2.29 \pm 0.49$.
All the quoted errors are at the 90\% confidence level.
 
We attempted to fit a model consisting of a power law with an
exponential cut-off to the prompt emission.  Such a fit did not
constrain $E_{\rm peak}$, the peak of the $\nu F(\nu)$\ spectrum.
However, it may be possible to use the results of \citet{zhang06b} to
estimate $E_{\rm peak}$\ for the prompt emission.  \citet{zhang06b} have
shown that due to the relatively narrow energy band of BAT, it is
often difficult to constrain $E_{\rm peak}$, even when $E_{\rm peak}$
is within the BAT energy range. These authors have found that the power-law photon index
$\Gamma$\ of a simple power law fit and $E_{\rm peak}$\ are well correlated with a relationship:
\begin{equation} 
\log E_{peak} = (2.76 \pm 0.07) - (3.61 \pm 0.26)\log\Gamma\ ,
\end{equation}\label{eqa}
under the assumption that $E_{\rm peak}$ is within the BAT energy
range.  \citet{saka06} have reached a similar conclusion and consistent result by combining
simulations and a study of bursts for which $E_{\rm peak}$ has been
determined.
We can use Equation~1 and the measured
$\Gamma = 1.61$, to find that $E_{\rm peak}\ =
103.1^{+34.0}_{-25.5}$\ keV. This value of $E_{\rm peak}$\ is
consistent with the majority of long GRBs detected by BAT.  However,
another possibility is that $E_{\rm peak}$\ is above the BAT energy
range ($> 150$\ keV), and $1.61$ is instead 
simply the low energy power law index $\alpha$.
According to the study of \citet{kane06}, a value of $\alpha\ = 1.61$\
is within the range of $\alpha$\ values found for BATSE bursts, although only 22 of
the 350 bursts in the BATSE sample (6.3\%) have $\alpha\ > 1.61$.  Since
we cannot exclude this second possibility, we will be conservative and quote only a lower
limit, $E_{\rm peak} > 77.6$\ keV (taking the lower limit on $E_{\rm peak}$\ as derived from
Equation~1).  As discussed in \S~\ref{flares},
$E_{\rm peak}$ is well constrained for the later flares.

\subsection{Swift-XRT}
\label{XRT}

\noindent The spacecraft slewed immediately to the BAT location of GRB
060714 and the {\em Swift} X-Ray Telescope (XRT) began observing the
burst at 15:13:39 UT ($T_0 + \sim99$\ s), first in Image mode
and then Photodiode (PD) mode.  The Image mode contains no spectral or
timing information and the PD mode data is severely corrupted by the
presence of two hot columns. Thus the first usable data is in
Windowed Timing (WT) mode beginning at $T_0 + 107$\ s (see \cite{hill04} 
for a description of XRT readout modes). Starting from $T_0 + $249~s all 
observations were carried out in Photon Counting (PC) mode.

The XRT data were processed with the  {\sc xrtdas} software (v.~1.7.1) developed 
at the ASI Science Data Center and included in the HEAsoft package 
(v.~6.0.4). Event files were calibrated and cleaned with standard 
filtering criteria with the {\sc xrtpipeline} task using the latest 
calibration files available in the {\em Swift} CALDB distributed 
by HEASARC. Event lists were selected in the 0.3--10.0~keV energy band and 
grades 0--12 for PC mode data and grades 0--2 for WT data were 
used in the analysis (see \cite{burro05} for a definition of XRT event 
grades).

The XRT PC image of the field clearly showed a bright fading X-ray object 
in the field \citep{gcn5321}.  The coordinates of the burst were 
determined by the XRT to be (J2000): 
RA:$15^\mathrm{h} 11^\mathrm{m} 26\fs5$ ($227\fdg8604$), 
Dec: $-6\arcdeg 33\arcmin 59\farcs3$ ($-6\fdg5665$) 
with a 90\% confidence error circle radius of $3.8\;$arcsec.

Events for temporal and spectral analysis of WT mode data were selected 
using a 40-pixel wide rectangular region centered on the afterglow.
Background events were extracted from a nearby source-free rectangular 
region of 40 pixel width. 
Data in PC mode during the first {\em Swift} orbit (from $T_0 + 249$\ s 
to $T_0 + 1610$\ s) were significantly affected by pile-up. By comparing 
the observed Point Spread Function (PSF) profile with the analytical model 
\citep{Moretti05}, we removed pile-up effects by excluding events 
within a 2 pixel radius circle centered on the afterglow position and using 
an outer radius of 20 pixels. From the second orbit, the afterglow count rate 
was below the XRT pile-up limit and events were extracted using a 10-pixel 
radius circle. The background for PC mode was estimated from a nearby 
source-free circular region of 50-pixel radius. 
Source count rates for temporal analysis were corrected for the fraction of 
PSF falling outside the event extraction regions and for pixels partially 
exposed.
Ancillary response files for the spectral analysis were generated with 
the {\sc xrtmkarf} task applying corrections for the PSF losses and 
pixel exposures. The latest response matrices (v.~008) available in the 
{\em Swift} CALDB were used and source spectra were binned to ensure a minimum 
of 20 counts per bin.

\begin{figure}
\plotone{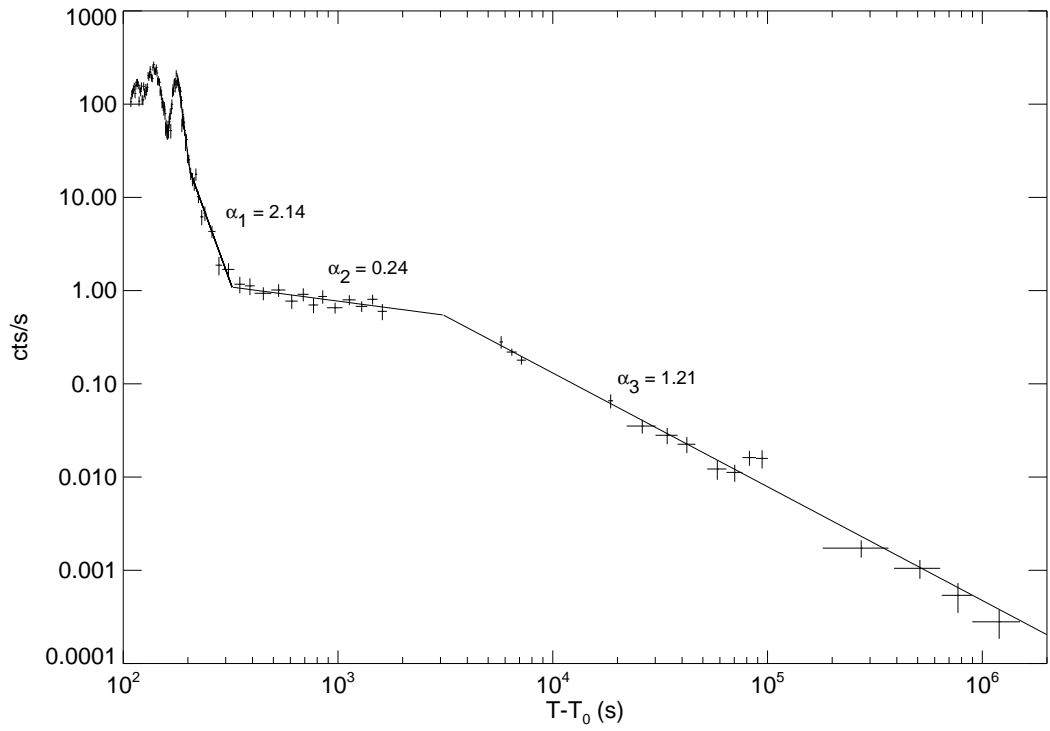}
\caption{The XRT light curve showing the early time flares and three
episodes of smoothly decaying afterglow emission.\label{fig2}}
\end{figure}

As discussed in detail in \S~\ref{sec3}, the XRT light curve \citep{gcn5321}
shown in Figures~\ref{fig4}, \ref{fig2},  and \ref{fig5} displays three flares 
during the first orbit peaking at about $T_0$\ + 115, 140 and 180 s after the 
BAT trigger. There is a steep decay after the end of the third XRT 
flare, with temporal power-law index $\alpha_1 = 2.14 \pm 0.13$,
followed by a much shallower decay starting at $t_{\rm break,1} \approx
330$\ s, which is probably the afterglow emission. The afterglow decay
from T$_0+324\;$s to T$_0+1.2\;$Ms can be fit with a broken power-law
with an initial decay slope of $\alpha_2 = 0.24 \pm 0.03$, a break at
$t_{\rm break,2} = 3.2^{+1.2}_{-0.7}\;$ks, and a post-break slope of
$\alpha_3 = 1.22 \pm 0.03$ \citep[we use here the notations of][]{nou06}.

A power-law fit to the 0.3--10 keV spectrum from T$_0+107\;$s to
T$_0+248\;$s (WT mode) gives a photon index of $\Gamma = 2.05 \pm 0.06$ and a
column density of $(2.26 \pm 0.20) \times 10^{21}\;{\rm cm}^{-2}$. The Galactic hydrogen column density in the direction of the
burst is $6.7 \times 10^{20}\;{\rm cm}^{-2}$.  An extrapolation backward in time
of the flat afterglow component ($\alpha_2$\ segment in Figure~\ref{fig2}) tells us that the
contribution of the underlying  afterglow during this period is negligible and can safely be ignored
in the spectral analysis.
We also fit  an 
absorbed single power-law model to the XRT 0.3--10 keV spectrum in PC mode (from T$_0+249\;$s to T$_0+1610\;$).  Here we 
found a photon index of $\Gamma = 2.2 \pm 0.2$ and a
column density of $(1.7 \pm 0.5) \times 10^{21}\;{\rm cm}^{-2}$. At later 
times, from T$_0+6096\;$s to T$_0+45416\;$s, the X--ray spectrum was well 
described by a single power-law model with photon index 
$\Gamma = 2.4 \pm 0.4$ and an absorbing column density of 
$(1.9 \pm 0.8) \times 10^{21}\;{\rm cm}^{-2}$.

\subsection{Swift-UVOT} 
\label{uvot}

\begin{figure}
\plotone{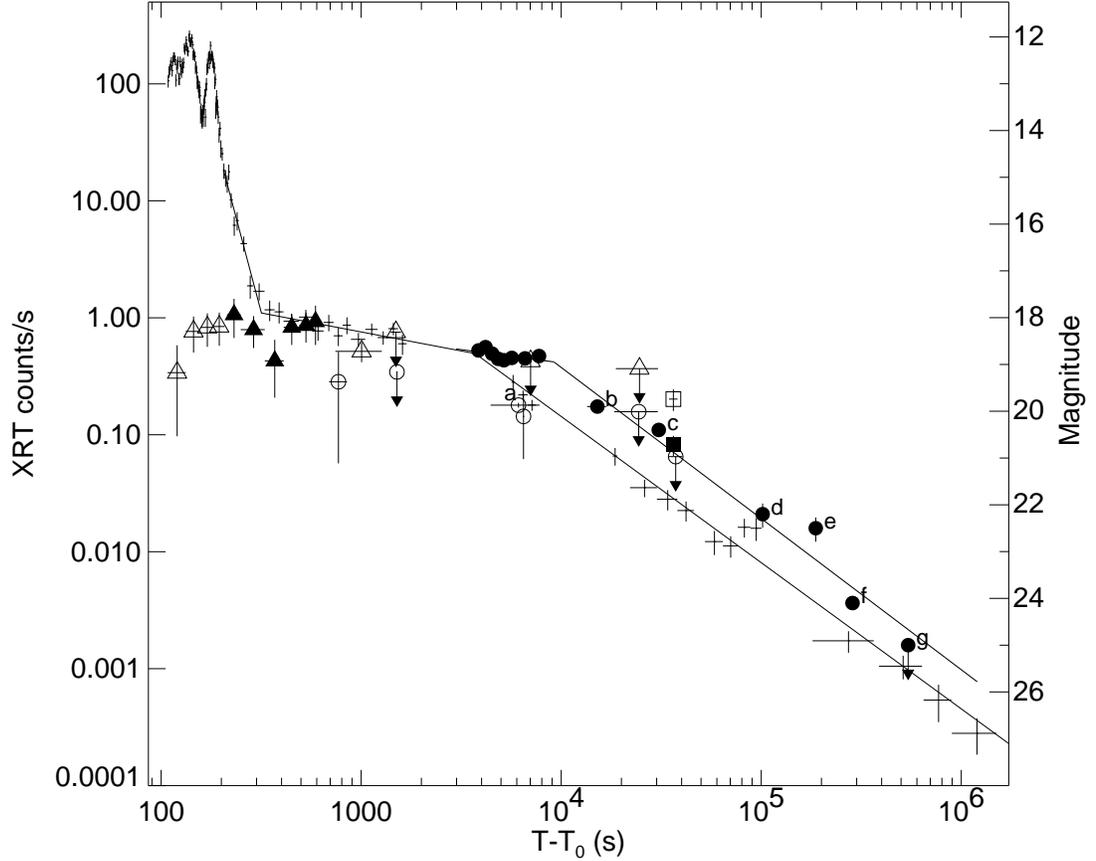}
\caption{The optical measurements (various symbols; right hand scale)
and X-ray count rate (crosses; left hand scale) for the late time
observations. The fits to the data discussed in the text are shown as
solid lines. The optical symbols are defined as follows: V: filled
triangles,  White: open triangles, B: open circles, R: filled circles, J: open square, I: filled
square. Credits:  I and J band: \citet{gcn5323}; unlabled R band:
\citet{gcn5434}; labeled points: a. \citet{gcn5434},
b. \citet{gcn5317}, c. \citet{gcn5320}, d. \citet{gcn5324},
e. \citet{gcn5336}, f. \citet{gcn5337}, g. \citet{gcn5355}. All other points are UVOT
measurements. U and W1 band
upper limits are omitted for clarity. All upper limits are $2\sigma$.\label{fig3}}
\end{figure}

\noindent The {\em Swift} Ultra Violet/Optical Telescope
\citep[UVOT;][]{romi05} began observing the field of GRB 060714 at
$T_0+90$\ s in the settling mode and obtained the first detection in the White filter  (160-650 nm) in the
exposure starting at $T_0+108$\ s \citep{gcn5357} . 
%An optical counterpart was detected in
%at a position 
%RA:$15^\mathrm{h} 11^\mathrm{m} 26\fs4$ ($227\fdg8601$), 
%Dec: $-6\arcdeg 33\arcmin 57\farcs6$ ($-6\fdg5662$)  (J2000) with a 90\% confidence interval of
%$0.6\;$arcsec. 
The position of the afterglow measured in the image from the initial 
exposure with the white filter is RA:$15^\mathrm{h} 11^\mathrm{m} 26\fs444$ ($227\fdg8602$), 
Dec: $-6\arcdeg 33\arcmin 58\farcs35$ ($-6\fdg5662$)  (J2000).
%RA  (J200) =  15 11 26.444
%Dec (J200) = -06 33 58.35
The uncertainty of this position is likely to be dominated by systematic
errors, which we estimate to be $\sim 0.5^{\prime\prime}$\ (90\% confidence radius) based
on residuals when matching UVOT sources in the image to stars in the USNO-B1.0 catalog \citep{monet03}.
UVOT used the standard sequence of exposures for observing gamma-ray bursts.
The sequence cycles through all seven lenticular filters with increasing exposure times
as the time from the trigger increases. The afterglow was strongly detected in the
White and V filters and weakly detected in the B filter.
Figure~\ref{fig3} shows the UVOT detections and
upper limits in the White, V and B bands, using UVOT CALDB version 20061116, and correcting for galactic extinction using the extinction curve of 
\citet{pei92} and the reddening from \citet{schl98}. 
The count rates in the White filter were converted to equivalent V
magnitudes using the ratio of the average count rates in White and V seen in the multiple
exposures between $T_0$+700 s. and $T_0$+1582 s.  Upper limits ($2\sigma$) in the U and W1 bands (omitted for clarity from Figure~\ref{fig3}) are
U $> 18.6$\ (667~s $<\ T_0\ < $ 835~s ), and  W1 $> 19.3$\ (643~s $<\ T_0\ <$ 811~s ). 
This lack of detection is consistent with the reported 
burst redshift of $z = 2.71$\ \citep[][See \S~\ref{other}]{jako06} which shifts the Lyman edge to 338 nm, cutting out much of the U and all of the W1 band.  The Lyman forest is also likely to reduce the flux in the U and B filters.

During the time period in which the X-ray light curve is showing flares followed by a steep decline, the optical light curve in the V band is essentially flat and at the same level as the afterglow.  This tells us that the flaring activity does not manifest itself in the optical and suggests that the afterglow component dominates the optical light curve even at early times.  This is consistent with what has been seen for other bursts with intense early flaring activity \citep[for example][]{roma06, guet07}, and for the unusual ``late plateau" of GRB~070110 \citep{troj07}.   At least during the time for which we have optical data, this burst seems to behave in a similar way to GRB~050401 \citep{ryko05} in which the optical and 
X-ray emission vary independently, but the prompt optical emission is consistent with a backward 
extrapolation of the later afterglow emission.

\subsection{Other observations} \label{other}

\noindent GRB 060714 was well observed by a large number of other
telescopes with detections beginning at T$_0 + 3860\;$s
\citep{gcn5434} and continuing until T$_0 + 3.31\;$days
\citep{gcn5337}.  A $2\sigma$\ upper limit of $R > 25.0$\ was obtained
at T$_0 + 6.28\;$days \citep{gcn5355}.  Figure~\ref{fig3} shows the
reported optical detections compared to the X-ray data.  During an
observation beginning $12\;$hr after the burst, with the FOcal
Reducer and low dispersion Spectrograph for the Very Large Telescope
of the European Southern Observatory,  [\citet{jako06} acquired a
spectrum showing numerous absorption features which correspond to
a redshift of $z = 2.711\ \pm 0.001$\ and a neutral hydrogen column density of $(6.3\ \pm 1.5)\ \times 10^{21}\ cm^{-2}$.  As seen in Figure~\ref{fig3}, the R-band
optical light curve can be well fit to a broken power-law decay with
an index $\alpha_{R,2} = 0.23 \pm 0.13$\ before T$_0 +\sim 10^4\;$s,
and $\alpha_{R,3} = 1.29 \pm 0.10$ after.  These values are very
similar to the decay constants for the X-ray afterglow, $\alpha_{X,2}
= 0.24 \pm 0.03$ and $\alpha_{X,3} = 1.22 \pm 0.03$, respectively.
However the break in the X-ray light curve occurs at least $3\;$ks
(corresponding to a factor of $\sim 2$ in time)
earlier than the break in the R-band light curve.
 
\section{Spectroscopy and Light Curves of the Flares}\label{sec3}
\label{flares}

\noindent There were a total of five flares detected in this burst as
seen in Figure~\ref{fig4}. The first two were seen only in the BAT:
the first was during the slew and the second during times when the XRT
was observing in Imaging and Photodiode mode and no detailed spectral
or temporal information is available. The third one was significant in
both BAT and XRT and the fourth and fifth were seen as strong flares
in the XRT, and weakly in the BAT.

	In the observer frame the flares lasted from $\sim 84$\ s to $\sim 210$\ s after the onset of burst emission (at $T-13.4$\ s).  Given the burst redshift of $z = 2.71$\ (a quite typical value for {\it Swift}), in the source frame the flares extend from $\sim 23$\ s to $\sim 56$\ s after the start of emission.  Although this means that the flares are relatively early as seen in the source frame, the timing of flares varies widely in the cosmological frame of GRBs, and there is no indication that absolute timing is a distinguishing characteristic of flares.  Although the timing analysis described in \S~\ref{spec} and discussed in \S~\ref{sec4} is done in the observer frame, the conclusions are all based on relative timing and do not depend on the cosmological frame chosen.

\subsection{Spectroscopy}\label{spec} 
 
\noindent All five flares were fit with a spectral model of a power
law with exponential cut-off:  $ F(E)\ =\ A(E/100\ {\rm keV})^{-\alpha} \exp[-E(2-\alpha)/E_{peak}] $, where $E$\ is the photon energy, $E_{peak}$\ is the peak energy of the $\nu F(\nu)$\ spectrum, $\alpha$\ is the photon index, and $A$\ is a normalization factor. The first two were fit to BAT alone, the
last three to BAT jointly with XRT.  For the fourth and fifth flares
inclusion of the BAT data did not significantly affect the fit.  The
bars at the bottom of Figure~\ref{fig4} indicate the time segments
used to fit each of the flares.  The results of the spectral fits are
shown in Table~\ref{tbl-3} and in the top two panels of Figure~\ref{fig6}.  
Although there is no evidence for a smooth power law decay underlying the flares, it is 
possible that they overlap each other temporally.  We have increased the error bars on the
flux and $E_{iso}$\ values in Table~\ref{tbl-3} to account for this overlap by extrapolating the
power-law fits (\S~\ref{timing}) to each flare down to zero and estimating the fraction of the flux
falling outside of the nominal start and stop times of the flare (for the upper limit), and the flux
possibly due to neighboring flares (for the lower limit).

In the fits to the last three flares an absorption component was included in the
model.  
The column densities were found to be (units $10^{21}$
cm$^{-2}$), respectively for flares 3-5, $1.91 \pm 0.43, 1.86 \pm 0.36, 1.67 \pm
0.22$, consistent within errors to a constant value.
 
One can see in Figure~\ref{fig6} that the peak
energies of the flares decrease with time and the power law indices
show a general softening of the spectra. Furthermore, the apparent linearity of
the plots
indicates a
connection (or at least a clear trend) between the five flaring
events. The time dependence of $E_{\rm peak}$ is well fit by a power
law, with index $-5.81 \pm 0.68$.  
Similarly, the time dependence of the power law index can be fit to a
power law with index $0.67 \pm 0.15$.

Using the redshift $z = 2.71$ for this burst, we extrapolate the total
isotropic equivalent radiated energy, $E_{\rm\gamma,iso}$ (in ergs) in the range
of 1--10$^4\;$keV, using the definition of \citet{amat02}.  For the flares, the extrapolation fixes the $E_{\rm
peak}$\ and $\alpha$\ values derived from the cut-off power law fits, and
uses a fixed high energy index $\beta = -10.0$.  For the prompt emission
we derive a lower limit to $E_{\rm\gamma,iso}$ from the lower limit to
$E_{\rm peak}$.  The third panel of Figure~\ref{fig6} shows
$E_{\rm\gamma,iso}$ as a function of time.  There is a general trend
towards lower total energy output of successive flares. Here a fit to
a temporal decay power law gives an index of $-1.72 \pm 0.46$. In
Figure~\ref{fig7} we plot $E_{\rm peak}$ against $E_{\rm\gamma,iso}$
for the flares and the prompt emission. This enables us to compare the
episodes of GRB~060714 to the $E_{\rm peak}-E_{\rm\gamma,iso}$
relationships 
found
by \citet{amat02} and parameterized by \citet{ghir04}.\footnote{Since we do not see a jet break in the light curve, we are unable to constrain the $E_{\rm peak}-E_{\rm\gamma}(\theta)$\ relation of \citet{ghir04}.} 
The prompt emission and the first two
flares fall on or very close to the relationships, while the last three
flares fall well below them.  This shows that $E_{\rm peak}$ is
falling with time more rapidly than the square of the total isotropic
equivalent energy emitted in the flares, $(E_{\rm\gamma,iso})^2$.

\subsection{Temporal Analysis} \label{timing}

In addition to the spectral properties of the flares, their temporal
properties can also help to tell us whether or not these events arise in the
external shock (i.e. are afterglow emission). 

In order to study the fine time structure of sub-peaks within the flares, we have derived a robust method to distinguish between a significant change in slope of the light curve (indicating the start or end of a new subpeak) and a statistical fluctuation.  Using one-second binned light curves for both the BAT and XRT data, we applied an iterative method to find each significant episode (subpeak) during the times of the flares .   In the first iteration, peaks were defined as local maxima (points higher than each of their nearest neighbor points) and valleys as local minima.  In successive iterations we culled the peaks by requiring that a significant peak be at least three standard deviations above the valleys on either side.  By this method, each point in the light curve was assigned to a specific interval:  either the rise or fall of a peak or to the periods of slow rise ($T < -13.4$\ s) or slow decay ($18.0\ {\rm s} < T < 70.2$\ s or $T > 105.4$\ s for BAT; and $T > 195.6$\ s for XRT).  We then fit each interval to a power law: log(R) vs. log(t), where R
is the count rate for either BAT or XRT (depending on which flare is
being analyzed) and t is the time since the burst trigger.  Finally we use these power law fits to define the actual start, apex and end of each subpeak (not restricted to light curve bin edges). The start of
each episode ($t_1$) is defined as the time at which the rising power
law segment of a peak crosses the falling power law segment of either
the preceding smooth decay or the previous peak.  Similarly the peak
time of each interval ($t_2$) is the time at which the rising power law
segment of a peak meets the falling power law of the same peak.  The
rise time ($\Delta t$) is thus defined as $\Delta t \equiv t_2 - t_1$,
while the time associated with each rise episode is $t \equiv (t_2 +
t_1)/2$. Thus
\begin{equation}\label{dt_t}
\frac{\Delta t}{t} \equiv \frac{2(t_2-t_1)}{(t_2+t_1)} = \frac{\Delta
  t}{(t_1+\Delta t/2)} = \frac{\Delta
  t}{(t_2-\Delta t/2)}\ .
\end{equation}
The conclusions drawn in \S~\ref{sec4} do not depend critically on the exact definitions of $t$\ or $\Delta t$.

The episodes so defined are shown graphically in Figure~\ref{fig5} and their main 
temporal properties in 
Table~\ref{tbl-1}.  The
temporal decay indices are derived in two ways, detailed in the table
caption.  The decay index is first calculated using the burst trigger
time $T_0$\ as the reference time ($\alpha_{\rm A}$), which is the
standard way that GRB decay indexes are calculated, and would be
appropriate for the flares if they were afterglow emission. However,
since the resulting values of the decay index $\alpha_{\rm A}$ are
very high, corresponding to a very steep decay which is very hard to
produce by the afterglow emission \citep{kumar00,NP03}, we also want
to explore the possibility that the flares arise from late time
sporadic activity of the central source. In this case the appropriate
reference time, $t_0$, would roughly correspond to the onset of the
individual flare or sub-flare whose decay rate we wish to quantify
($\alpha_{\rm B}$).
Both decay indices are shown in Table~\ref{tbl-1}, although values
quoted in Figure~\ref{fig2} are taken as the decay index $\alpha_{\rm
B}$.  Even taking the second definition, $\alpha_{\rm B}$, the decay
indexes of the flares are all very steep, and except for flare~4, much
steeper than either the decay of the emission just before the first
flare ($\alpha\ =\ 1.16 \pm 0.47$) or the decay immediately after the
fifth flare ($\alpha\ =\ 2.14 \pm 0.13$).  The decay slope immediately
after the last flare is still much steeper than the afterglow
beginning at $T\ +\ \sim 320$\ s, so it is likely part of the prompt
emission as well. We note, however, that $\alpha_{\rm B}$ does not
exceed $2+\beta$ (within the statistical uncertainty, where $F_\nu
\propto \nu^{-\beta}(t-t_0)^{-\alpha}$), which is the steepest decay
allowed by the `high latitude' emission \citep{kumar00}, and thus the
decay of the flares and sub-flares is consistent with the expectations
for late time intermittent activity of the central source.

\section{Discussion}\label{sec4}

\subsection{The External Shock}\label{extshock}

The temporal properties of the flares provide strong evidence against
an external shock (i.e. afterglow) origin for
them. Figure~\ref{dtt_dff} shows the fractional increase in flux,
$\Delta F/F$, versus the ratio of the rise time ($\Delta t$) to the
peak of each flare or sub-flare and the time ($t$) from the GRB
trigger, $\Delta t/t$ (see Eq.~\ref{dt_t}). It can clearly be seen
that large amplitude variations in the flux, $\Delta F/F \gtrsim 1$,
occur on very short time scales, $\Delta t/t \ll 1$. This basically
rules out an external shock origin for the flares \citep[see,
e.g.,][]{ioka05,Nakar06,lazz07}.  We also note that the flares are very
different from the smooth, late time tail emission in the 20-100~keV band 
which was observed to follow many of the bursts detected by the Burst and
Transient Source Experiment (BATSE) \citep{conn02}.  The tail emission is 
temporally much smoother than the flares of GRB~060714, and at a much lower
level (as a ratio to the peak of the prompt emission).  Also \citet{gibl02} identify
a subset of the BATSE bursts that have high-energy decay emission consistent
with forward external shocks.  Although it is below detectability for BAT, it is 
possible that the tail emission is seen after the last flare in the $\alpha_1$\ = 2.14
decay segment (Figure~\ref{fig2}).  This decay index is consistent with the 
average temporal decay index reported for BATSE tail emission, $-2.03\ \pm 0.51$\
\citep{gibl02}.

The major possible sources of variability in the afterglow light
curves (i.e. in the emission from the external shock) are: (i) a
variable external density \citep{WL00,Lazzati02,NPG03}, (ii) a
``patchy shell'' i.e. angular inhomogeneity within the outflow
\citep{KP00b,NPG03}, and (iii) ``refreshed shocks'' i.e. relatively
slow shells that were ejected from the central source toward the end
of the prompt emission and catch up with the afterglow shock as the
latter decelerates to a Lorentz factor slightly lower than that of the
shells \citep{rees98,KP00a}. Such a sharp ($\Delta t/t \ll 1$) large
amplitude ($\Delta F/F \gtrsim 1$) rise in the observed flux, as we
find for GRB~060714 (see Figure~\ref{dtt_dff}), cannot be caused by a
sudden increase in the external density \citep{NG06}. In addition, a
``patchy shell'' produces $\Delta t/t \sim 1$ \citep{NO04}, and cannot
account for the observed $\Delta t/t \ll 1$, since new `bright spots'
in the outflow become visible (i.e. enter the observed region of angle
$\sim 1/\gamma$ around the line of sight) gradually, on the dynamical
time ($\Delta t \sim t$). Finally, ``refreshed shocks'' produce
$\Delta t/t \sim 1$ before the jet break time, when the rise time
$\Delta t$ is dominated by the angular time $t_\theta \approx
R/2c\gamma^2$. After the jet break time, the angular time is $t_\theta
\approx R\theta_j^2/2$ (where $\theta_j$ is the half-opening angle of
the jet), and it soon becomes smaller than the radial time $t_r
\approx R/10c\gamma^2$ \citep[since $\theta_j$ remains close to its
initial value, $\theta_0$, as long as the jet is
relativistic;][]{Granot06}, so that the rise time $\Delta t$ becomes
dominated by the radial time \citep{GNP03}.  However, even in this
case $\Delta t/t \gtrsim 0.2$, which cannot explain the values of
$\Delta t/t < 0.1$ and in some cases even as low as $\Delta t/t
\lesssim 10^{-2}$, that we obtain in our analysis.

\subsection{Other Possible Causes for the Flares}

Now that we have effectively excluded an external shock origin for the
flares, we explore other possible explanations. These involve sporadic
late time dissipation events which are a result of either (i) early
time activity of the central source on a time scale comparable to the
duration of the prompt GRB emission, or (ii) late time intermittent
activity of the central source, which is rather directly reflected in
the observed times of the flares. Both types of models may in principal
be applicable both for the prompt GRB emission and for the flares, due
to their roughly similar observed properties. Thus we also wish to
address the similarities and differences in the observed temporal and
spectral properties of the flares and the prompt emission.

Models in the literature that can be included in the first of the two
above classes include both emission powered by sporadic magnetic
dissipation events within the outflow, possibly induced by its
interaction with the external medium \citep{LB02,gian06,Thompson06}, and late
time internal shocks between shells with a small relative velocity
\citep[e.g.,][]{Barraud05,burro07}. In the former models the
temporal and spectral properties of the emission have not yet been worked
out in detail, so direct comparison to observations is not possible
at this stage. We now briefly address the latter model.

\subsection{Internal Shocks with Small Contrast in the Lorentz Factor}
\label{ISa}

In this scenario the difference in the Lorentz factors of the
colliding shells, $\Delta\gamma$, is much smaller than their typical
Lorentz factor $\gamma$: $\Delta\gamma \ll \gamma$.  For convenience,
we provide most of the relevant results here, while a detailed
derivation of these results as well as some relevant (but somewhat
more technical) discussion is provided in
Appendix~\ref{low_contrast_IS}. In this picture, later collisions
correspond to a smaller $\Delta\gamma/\gamma$, occuring at an observed
time $t_{\rm flare}$ and with a duration $\Delta t_{\rm flare}$ which
satisfy
\begin{equation}
t_{\rm flare} \sim t_{\rm ej} + \frac{\gamma}{\Delta\gamma}
\Delta t_{\rm ej}
\quad,\quad
\Delta t_{\rm flare} \sim
\frac{\gamma}{\Delta\gamma}\Delta t_{\rm ej} 
\sim t_{\rm flare} - t_{\rm ej}\ ,
\end{equation} 
where the first and second shells are ejected at times $t_{\rm ej}$
and $t_{\rm ej} + \Delta t_{\rm ej}$ with Lorentz factors $\gamma$ and
$\gamma + \Delta\gamma$, respectively. This makes it hard to account
for the short time scale variability $\Delta t/t \ll 1$, unless the
colliding shells were ejected from the source at a time $t_{\rm ej}$
after the GRB trigger that is much closer to the observed time of the
flare $t_{\rm flare}$ than to the GRB trigger $t = 0$ (since $\Delta
t/t =\Delta t_{\rm flare}/t_{\rm flare} \sim 1 - t_{\rm ej}/t_{\rm
flare}$). However, this corresponds to models of class (ii) above
(where the central source is active at late times, close to the time
of the observed flares), rather than class (i). \citet{lazz07} come to a 
similar conclusion.

In the latter case, which corresponds to class (ii), one can in
principal account for the temporal properties of the flares. We also
wish to examine whether their spectral properties and energetics can
naturally be reproduced. In this picture the shells collide at a radius
\begin{equation}
R_{\rm IS} \approx 
\frac{\gamma}{\Delta\gamma}\,\gamma^2c\Delta t_{\rm ej}\ , 
\end{equation}
(the subscript `IS' is for internal shocks) which is larger by a
 factor of $\sim \gamma/\Delta\gamma \gg 1$ (for the same average
 $\gamma$) compared to internal shocks with a reasonably large
 contrast in the Lorentz factor, $\Delta\gamma \gtrsim \gamma$. For a
 reasonably large contrast in the Lorentz factor ($\Delta\gamma/\gamma
 \gtrsim 1$) the peak of the $\nu F(\nu)$\ spectrum, $E_{\rm peak}$,
 typically corresponds to $h\nu_m$ where $\nu_m$ is the synchrotron
 frequency of the relativistic electrons with the minimal random
 Lorentz factor in the power law distribution of energies. However, for
 low contrast internal shocks ($\Delta\gamma/\gamma \ll 1$),
\begin{equation}
\nu_m \propto \left(\frac{\Delta\gamma}{\gamma}\right)^5 
L_{\rm iso}^{1/2}R_{\rm IS}^{-1} \propto 
\left(\frac{\Delta\gamma}{\gamma}\right)^6 L_{\rm iso}^{1/2}
\gamma^{-2}(\Delta t_{\rm ej})^{-1}\ ,
\end{equation}
where $L_{\rm iso}$ is the isotropic equivalent kinetic luminosity of
the outflow, under the standard assumptions that the fractions of the
internal energy behind the shock in the relativistic electrons
($\epsilon_e$) and in the magnetic field ($\epsilon_B$) are
constant. That is, $\nu_m$ decreases very rapidly for a small contrast
$\Delta\gamma/\gamma$. On the other hand, the cooling break frequency
$\nu_c$ increases as $\Delta\gamma/\gamma$ decreases,
\begin{equation}
\nu_c \propto \left(\frac{\Delta\gamma}{\gamma}\right)^{-3} 
L_{\rm iso}^{-3/2}R_{\rm IS}\gamma^6 \propto 
\left(\frac{\Delta\gamma}{\gamma}\right)^{-4} L_{\rm iso}^{-3/2}
\gamma^8\Delta t_{\rm ej}\ .
\end{equation}
Therefore, for $\Delta\gamma/\gamma \ll 1$, $\nu_c > \nu_m$ and there
 is slow cooling, i.e. $E_{\rm peak} = h\nu_c$ rather than $h\nu_m$.
 
For a reasonable value of the magnetic field that is advected with the
 outflow from the central source, such a field would dominate over a
 sub-equipartition shock-generated field in the shocked regions of the
 colliding shells (see Appendix~\ref{low_contrast_IS} for details). In
 that case $\nu_m/\nu_c$ scales ``only'' as $(\Delta\gamma/\gamma)^6$,
 instead of $(\Delta\gamma/\gamma)^{10}$ for a field that is a
 constant fraction of the equipartition value, that was assumed above.
Since this is still a very high power of $\Delta\gamma/\gamma$, this
 does not change the main conclusions.

 The very strong dependence of $\nu_m/\nu_c$ on $\Delta\gamma/\gamma$
 further decreases the radiative efficiency, which is already low for
 $\Delta\gamma/\gamma \ll 1$, since the fraction of the total energy
 of the colliding shells that is converted into internal energy (out
 of which the fraction that goes into relativistic electrons,
 $\epsilon_e$, may be mostly radiated away for $\nu_c < \nu_m$ but not
 for $\nu_c > \nu_m$) is given by
\begin{equation}
\epsilon \approx \frac{x}{2(1+x)^2}
\left(\frac{\Delta\gamma}{\gamma}\right)^2 
\leq \frac{1}{8}\left(\frac{\Delta\gamma}{\gamma}\right)^2\ ,
\end{equation}
where $x$ is the ratio of the rest masses of the two colliding shells
(for a fixed contrast of the Lorentz factor, $\Delta\gamma/\gamma$,
the efficiency $\epsilon$ is maximal for equal mass shells, $x = 1$).
For GRB~060714 the combined $E_{\rm\gamma,iso}$ of the five flares is
comparable to that of the prompt emission, and that of individual
flares is at most one order of magnitude smaller. This suggests that the
kinetic energy that remains in the colliding shells that produce the
flares in this scenario is much larger than that of the original
shells that produced the prompt emission. This would result in very
significant episodic energy injection into the afterglow shock,
i.e. ``refreshed shocks'' which are inconsistent with the temporal properties of 
GRB~060714 (and may prove problematic for most {\it Swift} GRBs with
X-ray flares). Furthermore, since the total
$E_{\gamma,iso}$ of the flares is $\sim 10^{52.5}\;$erg, such an extremely
inefficient emission would imply a huge remaining isotropic equivalent
kinetic energy, in excess of $10^{55.5}\;$erg for $\Delta\gamma/\gamma
\lesssim 0.1$, which would imply a very large total kinetic energy
($\gtrsim 10^{53}\;$erg), even when corrected for the fractional solid
angle occupied by the jet ($f_b \gtrsim 10^{-2.5}$), given that there is
no sign of a jet break at least until $\sim 10^6\;$s.

\subsection{Internal Shocks with Reasonably Large Contrast in 
the Lorentz Factor}\label{ISb}

We now examine the more popular version of the internal shocks model,
where the difference in the Lorentz factor of the colliding shells,
$\Delta\gamma$, is of the order of their average Lorentz factor
$\gamma$, i.e. $\Delta\gamma \gtrsim \gamma$). In this case the
observed time of the flares reflects the time in which the colliding
relativistic shells were ejected from the source \citep{KPS97,NP02},
thus requiring intermittent late time activity of the central
source \citep{burro05b}.  Such a sporadic late time activity of the central source may
arise \citep[e.g.,][]{per06} by infall of material into the central
black hole from instabilities in an accretion disk, or by late time
fall-back of material in the collapsar model \citep[e.g.,][]{MWH01}.
\citet{fan05} show that this model is consistent with X-ray flares in a 
number of different bursts.

In this scenario the radiative efficiency may be reasonably high (but
still typically $\lesssim 10\%$), and $E_{\rm peak}$ is usually given
by $h\nu_m$. All of the equations of \S~\ref{ISa} still remain valid,
up to a factor of order unity, under the substitution
$\Delta\gamma/\gamma = 1$. Thus $\Delta t_{\rm flare} \sim \Delta
t_{\rm ej}$ and $E_{\rm peak} \propto L_{\rm
iso}^{1/2}\gamma^{-2}(\Delta t_{\rm ej})^{-1} \propto L_{\rm
iso}^{1/2}\gamma^{-2}(\Delta t_{\rm flare})^{-1} $.  Figure~\ref{fig8}
shows that the rise time $\Delta t \sim \Delta t_{\rm flare}$
increases with time (by about two orders of magnitude), while $\Delta
t/t$ increases by a somewhat smaller factor (but still a factor of
$\sim 30$ over less than a factor of 3 in time). This alone can
roughly account for the decrease in $E_{\rm peak}$ (see upper
panel of Figure~\ref{fig6}), which decreases by about two orders of
magnitude, similar to the increase in $\Delta t_{\rm flare}$. Since
for each flare $L_{\rm iso} \sim E_{\rm\gamma,iso}/\Delta t_{\rm
flare}$, and $E_{\rm\gamma,iso}$ of the individual flares decreased by
about one order of magnitude, we can infer that $L_{\rm iso}$
decreased by about three orders of magnitude, which requires $\gamma$
to decrease by a factor of $\sim 10^{3/4} \sim 5-6$. Since $R_{\rm IS}
\sim \gamma^2c\Delta t_{\rm ej} \sim \gamma^2c\Delta t_{\rm flare}$,
this implies a modest increase in $R_{\rm IS}$ by a factor of $\sim 3$
(since $\gamma^2$ decreased by a factor of $\sim 30$ while $\Delta
t_{\rm flare}$ increased by a factor of $\sim 100$).

\subsection{Possible Relation to Models for the Late Time Central
  Source Activity}\label{relation}

The sharpness of the peak features can be seen in Figure~\ref{fig8} in
which the rise time $\Delta t$\ and the ratio $\Delta t / t$\ are
plotted against the time since the trigger.  There is a tendency
for peaks to become less sharp (increasing $\Delta t$) with time.
This trend is consistent with the viscous disk evolution model
discussed by \citet{perna06}, in which both the accretion time scale
$\Delta t$\ and the arrival time $t$\ depend on the radial location of
the accreting material within the accretion disk.  \citet{koce07} show that
the correlation of broader pulse durations with time holds in general even among
flares from different GRBs, supporting the idea that later flares are coming from 
larger radius shells. The lower
panel in Figure~\ref{fig8} shows that even for the late flares with
longer rise times, that $\Delta t / t$\ remains very small, $\lesssim
0.1$\ for all peaks, although $\Delta t / t$ tends to increase with
time (it reaches values $< 10^{-2}$ at early times).  

The 
distribution of $\Delta t / t$ values for GRB~060714 can be compared
to the statistical samples of flares presented by \citet{chin07} and \citet{burro07}.
We see that most of the points in our 
$\Delta t / t$\ distribution are consistent with the \citet{chin07} sample, while the three points with 
$\Delta t / t < 0.01$\ lie outside the distribution.  This is explained by the
difference between our definition of $\Delta t$\ (rise time) and that used by \citet{chin07} ($\sigma$\ of a Gaussian fit to the peak).  Although these measures are of the same order of magnitude, our definition
leads to very small values of  $\Delta t / t$\ for those peaks with fast rises and slower decays, and in
general a broader distribution than seen by  \citet{chin07}.
Given the asymmetric shape of many of the peaks, the
rise time may be a truer measure of the variability, since the width can be
skewed by a slow decay. 
\citet{burro07} show $\Delta t_{rise} / t$, where $\Delta t_{rise}$\ is defined as the time between when
the power law crosses the underlying afterglow and the peak time, a definition which would lead to 
somewhat larger values of $\Delta t_{rise} / t$\ than ours.  Their distribution, however, has a median value $\sim 10$\ times larger than ours, indicating that the time structure in GRB~060714 is unusually sharp compared to most other flares.  The analysis of \citet{koce07} also shows that
GRB~060714 has shorter flare durations and rise times than most bursts.
Alternatively, it could be that we find significant structure on smaller time scales than
in other works because we are analyzing brighter flares with better
photon statistics.  We are also looking at sub-structure in the flares,
as is often done with the prompt emission, in which significant temporal structure is
usually found down to the smallest time scale that can be measured (which
is limited either by photon statistics or the temporal resolution of the
instrument).

The successive flares in GRB~060714 are each spectrally softer than
the preceding flare, as seen in the top panel of
Figure~\ref{fig6}. The spectral softening of the flares is supportive
of models in which multiple masses accrete onto a central black hole
at successively later times.  
\citet{king05} discuss a model where the collapse of a rapidly
rotating stellar core leads to fragmentation and the formation of
multiple compact objects which accrete onto the central black hole on
time scales which depend on the orbital radii and fragment masses.
\citet{king05} and \citet{burro05b} suggest that successively later
accretion events occur in cleaner, lower density environments, since
earlier jets would have excavated channels through the progenitor
star. This suggests that later outflows would have lower baryon
loading and hence higher $\gamma$.  \citet{king05} also suggest that
tidal effects may smooth later accretion events, lengthening $\Delta
t_{\rm ej}$.  Both increasing $\gamma$\ and increasing $\Delta t_{\rm
ej}$ would lead to a reduction in $E_{\rm peak}$ as a function of time
(see \S~\ref{ISb}).  Note that $E_{\rm peak}$\ for the prompt emission
is slightly higher than $E_{\rm peak}$\ for the first flare, which
would be expected since the initial collapse should have the lowest
bulk Lorentz factor. However, as demonstrated in \S~\ref{ISb}, while
Figure~\ref{fig8} indeed shows that $\Delta t \sim \Delta t_{\rm
flare}$ increases with time, we find that $\gamma$ should have {\it
decreased} in time (by a factor of $\sim 5-6$) rather than increased
with time (for the standard internal shocks model) as expected in the
above model.

As can be seen in the third panel of Figure~\ref{fig6},
$E_{\rm\gamma,iso}$ decreases with time roughly as $\sim t^{-1.7}$,
and $L_{\rm iso}$ decreases roughly as $t^{-3}$. This is a steeper
decline than the expected late-time accretion rate due to fallback in
the collapsar model, $\dot{M}_{\rm acc} \propto t^{-5/3}$
\citep[e.g.,][]{MWH01}, and in fact drops roughly as $\dot{M}_{\rm
acc}^2$, suggestive of a relativistic outflow powered by neutrino-anti
neutrino annihilation, where $L_{\rm iso} \propto L_\nu^2 \propto
\dot{M}_{\rm acc}^2$ (this is, of course, highly speculative at this
stage). Furthermore, such a steep decay of $L_{\rm iso}$ is much
steeper than $\propto t^q$ with $q > -1$ that is required in order to
account for the flat decay phase \citep{nou06}. Although the latter
requires a temporally smooth outflow, and the flares require 
intermittent source activity, it is hard to see how the overall
temporal decay of such a smooth component and a variable component
would be that different.

\subsection{Similarities and Differences between the Flares and the
  Prompt Emission}

\begin{figure}
\plotone{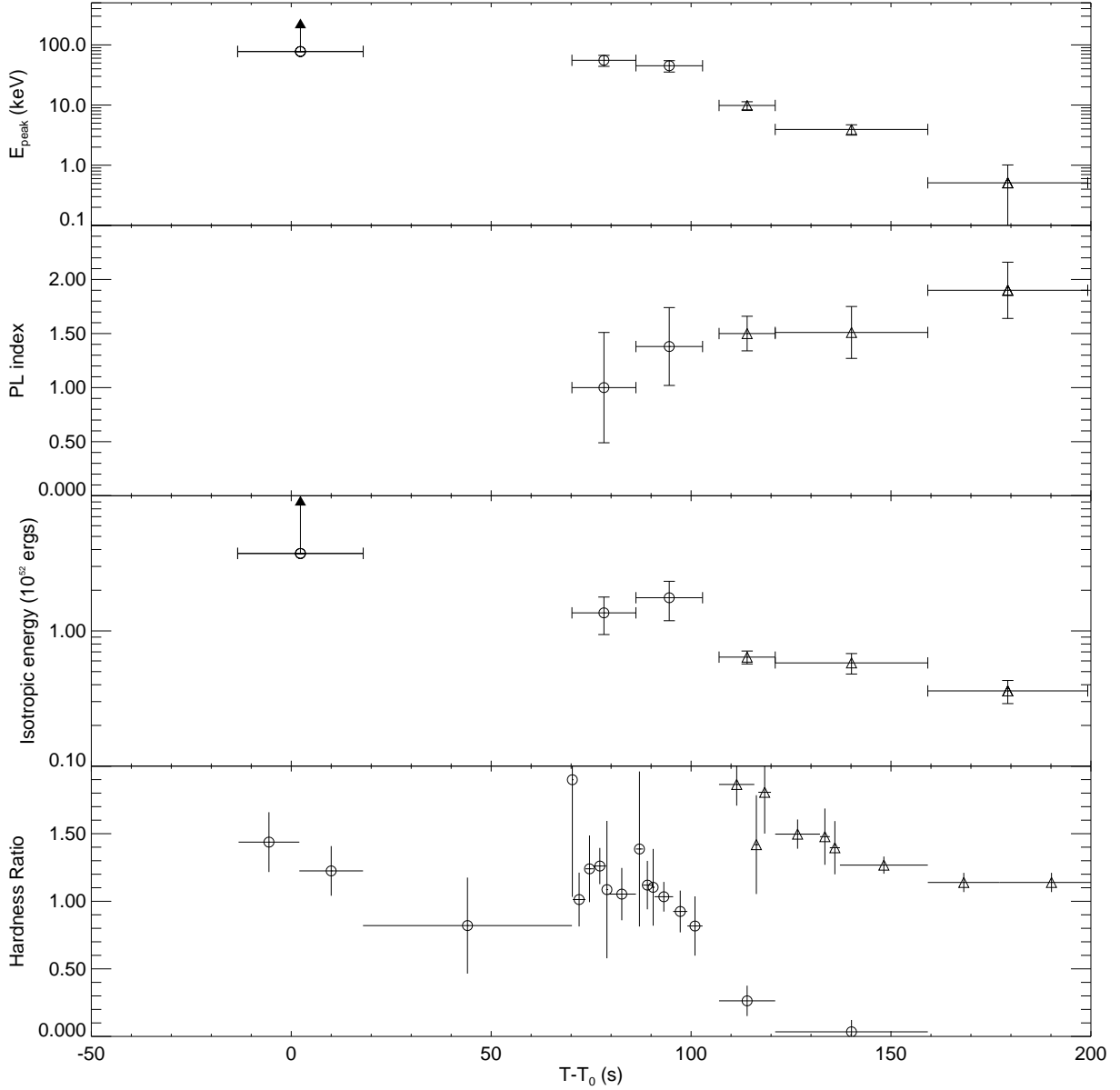}
\caption{\label{fig6} \small
The evolution of spectral fit and energetic properties of the flares.
The fits are to: circles -- BAT alone, triangles -- BAT and XRT
jointly.  We were unable to fit a cut-off power law to the prompt
emission so the points for the prompt emission show
lower limits on $E_{\rm peak}$\ and $E_{\rm iso}$.  The top panel
shows how $E_{\rm peak}$\ (shown in the observer frame)
changes from the prompt emission through the
five flares.  The second panel shows the evolution of the power-law
index ($\alpha$\ in the cut-off power law fit) across the flares.  Both of the two top plots clearly show a
hard to soft spectral evolution as the flares progress. The third
panel shows the time evolution of the isotropic radiated energy over
the 1-$10^4$\ keV energy range, indicating the the flares become
progressively less energetic.  The data plotted in the top three panels is also given in 
Table~\ref{tbl-3}. The bottom panel shows hardness ratios
for the burst.  For the prompt emission and the first four flares
the flux ratio S(50--100 keV)/S(25--50 keV) is shown as circles.  (The hardness
ratio for the last flare is consistent with zero.) For
the last three flares, we also show the flux ratio
S(1.5--10 keV)/S(0.3--1.5 keV) (triangles).}
\end{figure}

\begin{figure}
\plotone{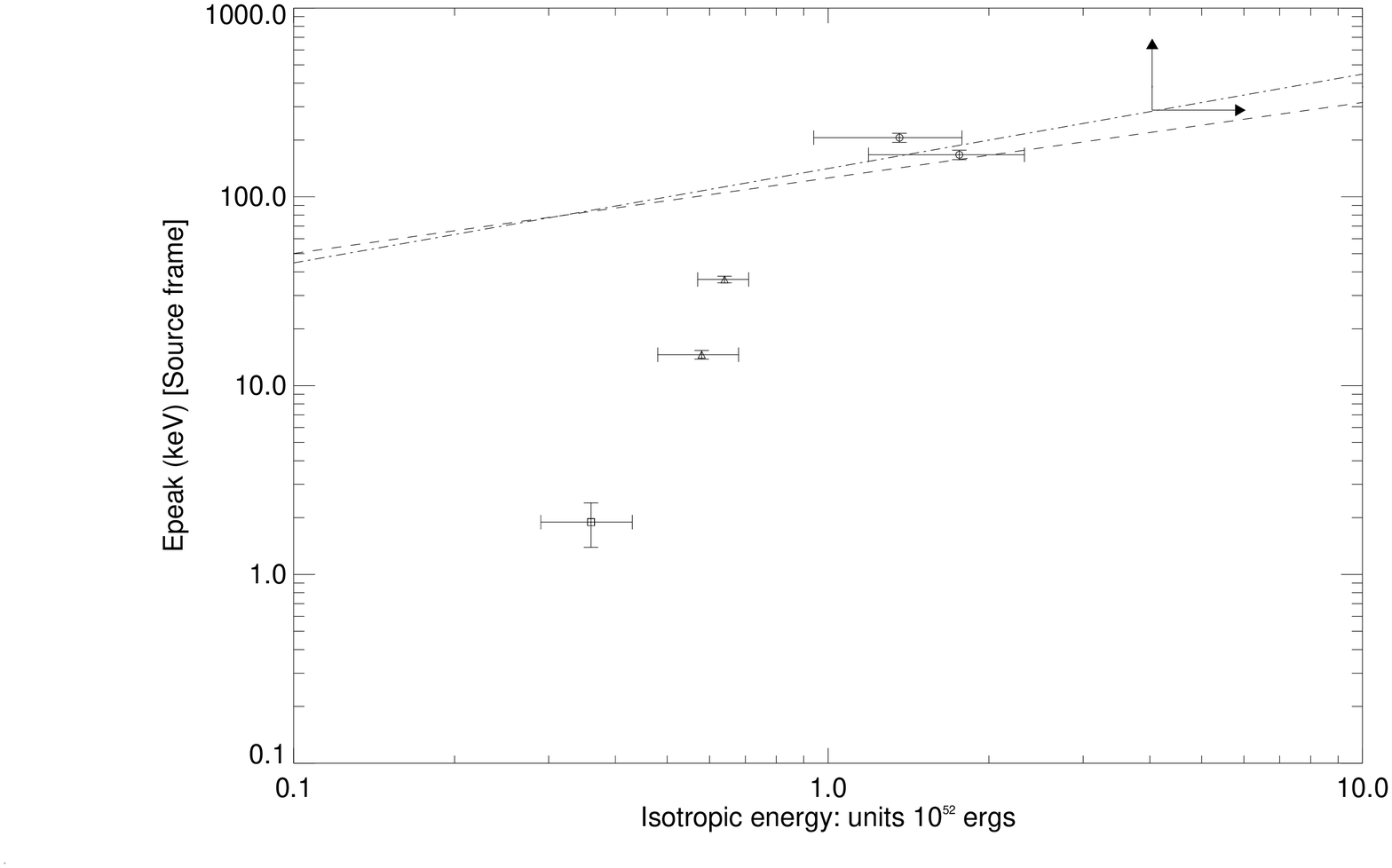}
\caption{The peak energy is plotted against the isotropic energy over
the 1-$10^4$\ keV energy range  \citep[the same energy range used by][]{amat02}. The dashed line is the fit to
$E_{peak}$-$E_{iso}$ derived by \citet{ghir04} and the dot-dashed line
is the fit derived by \citet{amat02}. The time ordering of the flares
goes monotonically from highest $E_{peak}$ to lowest.\ The prompt
emission and the first two flares detected fall on the Amati
relation while the last three flares fall below \label{fig7}}
\end{figure}

\begin{figure}
\plotone{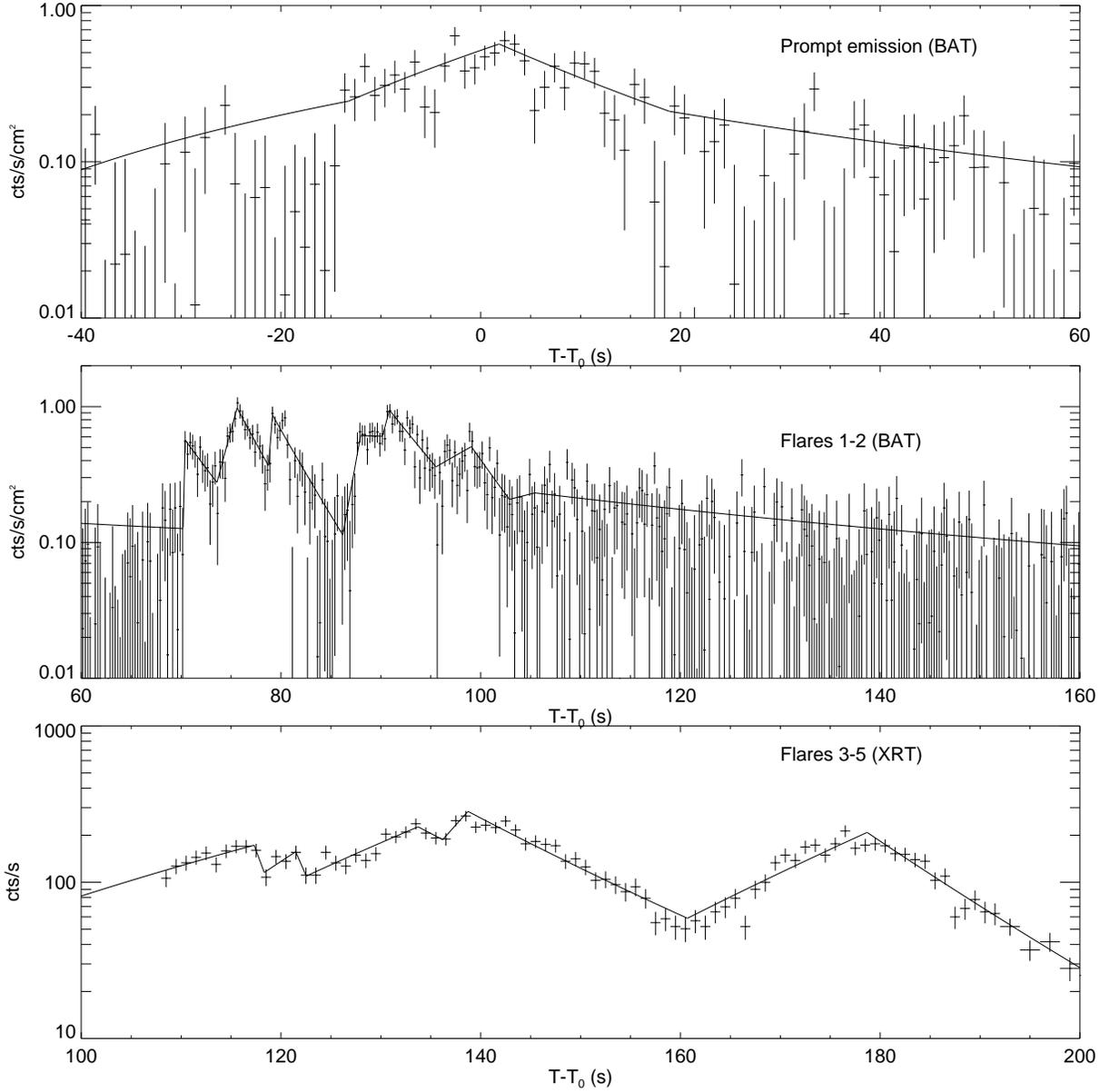}
\caption{The prompt emission (top), the BAT-detected flares (middle)
and the XRT-detected flares (bottom) are shown on the same time scale.
The solid lines on the plot show the best power law fit to each individual segment of the light curve and the changes in slope indicate the start, apex and end of each subpeak.  The method for defining the intervals is given in the text and decay constants are listed in Table~\ref{tbl-1}.   One can note the sharp temporal
features (subpeaks) of the flares.  It is also clear that the X-ray flares are longer than the BAT
flares, a feature also seen in \citet{roma06}, and consistent with previous results showing that
the duration of pulses in prompt emission are longer at low energy than at high energy. \label{fig5}}
\end{figure}

\begin{figure}
\plotone{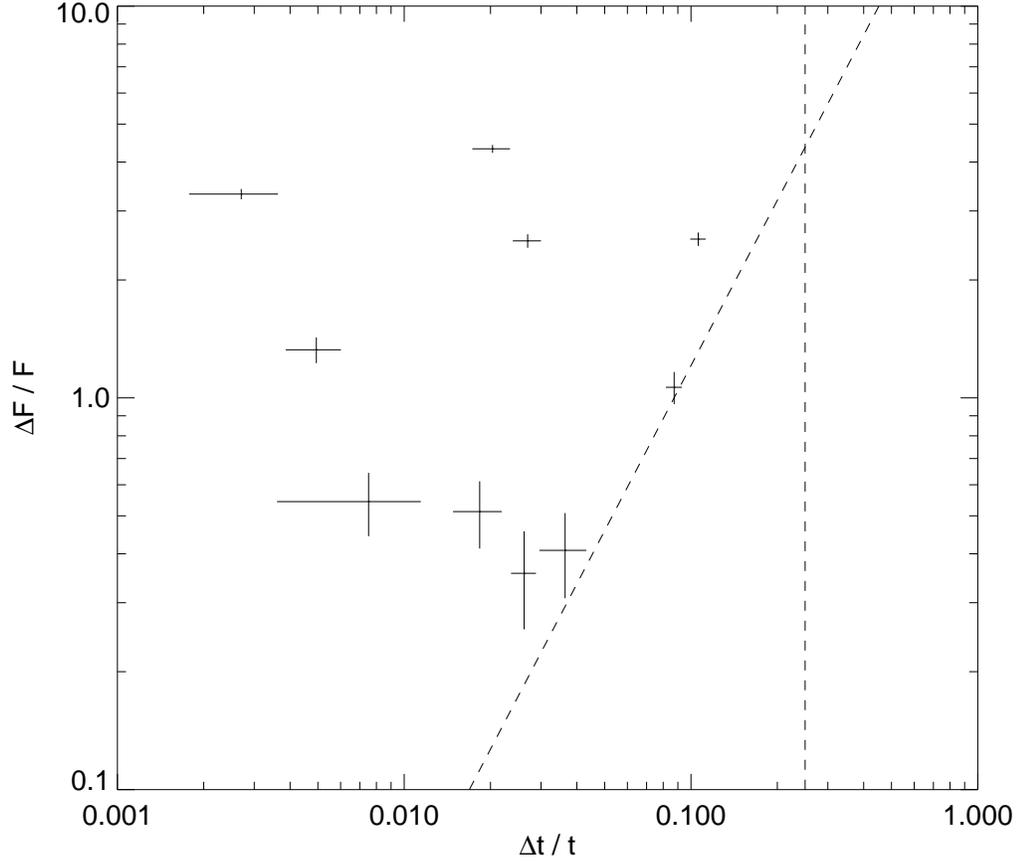}
\caption{Plot of $\Delta t / t$\ vs. $\Delta F / F$. The flux ratio
$\Delta F / F$\ is derived as $\Delta F = (F(t_2) - F(t_1))$\ and $F =
F(t_1)$, where $F(t_1)$\ and $F(t_2)$\ are the flux at the start and
top of each peak, respectively. In this notation $\Delta F / F$\ = 1
means a doubling of the flux. This figure shows that for all peaks
$\Delta F / F \sim 1$\ while $\Delta t / t \ll 1$.  The area to the
right of the dashed lines indicates the kinematically allowed region
for afterglow variability derived by \citet{ioka05}.  The vertical line indicates that 
refreshed shocks cannot make a bump with $\Delta t < t/4$. \citet{ioka05} also argue
that ambient density fluctuations cannot make a bump in afterglow
light curves larger than the limit indicated by the diagonal line.\label{dtt_dff}}
\end{figure}

\begin{figure}
\plotone{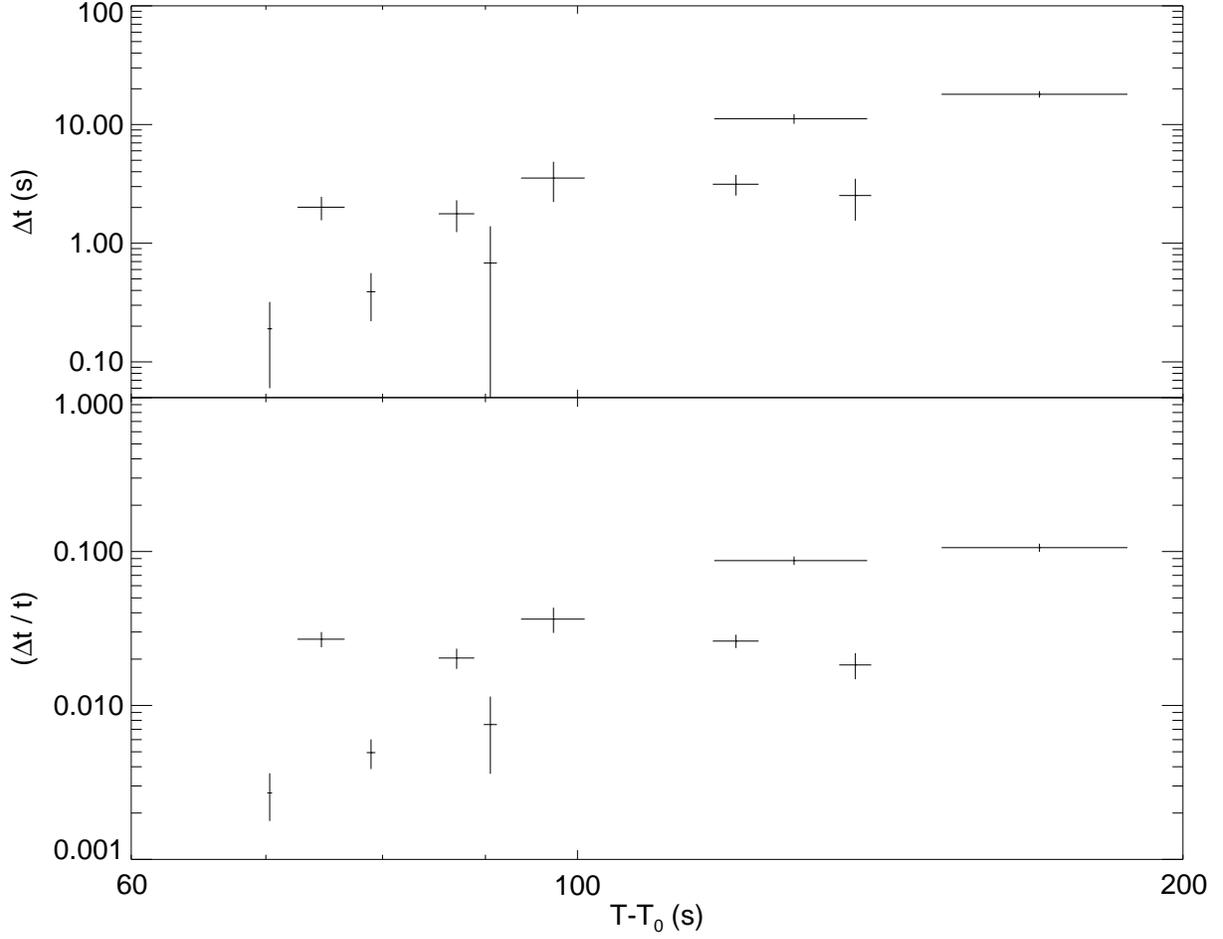}
\caption{This plot shows $\Delta t$ (rise time to the peak) and
$\Delta t / t$\ vs. time since the GRB trigger for each peak in each
of the flares. Note that the plot is log-log. The quantities $\Delta t$\ and $t$\ are defined in the
text.  Although there is some scatter there is a general trend for
both $\Delta t$\ and $\Delta t / t$\ to increase with t for the
flares.  A fit to $\Delta t$\ vs. time (top plot) gives a slope of
$4.0 \pm 0.9$.  
It is also quite clear that for all peaks in the flares, that $\Delta
t / t \ll 1$\label{fig8}}
\end{figure}

A striking feature of the spectroscopic results is that not only is
$E_{\rm peak}$ declining with each flare, but the low energy spectral
index $\alpha$ is increasing (second panel of Figure~\ref{fig6}),
meaning that the flares are becoming successively softer with time.
This is consistent with the third panel of Figure~\ref{fig6} which
shows the the total isotropic energy is decreasing with time.
Hard-to-soft spectral evolution is commonly seen in the prompt
emission of GRBs \citep{band98,nemir94}, and the bottom panel of
Figure~\ref{fig6} shows this trend very clearly for the period before
the start of the first flare (first three points).  For the flares,
not only is there a softening from flare to flare, but {\em within}
each of the first four flares there is a tendency for the spectrum to
soften as well.  That this well-established feature of the prompt
emission of GRBs is seen within and across the flares 
favors similar origins for the flares and the prompt emission.  
\citet{butl07} have shown for a sample of 27 {\em Swift}-detected GRB
flares, that flares exhibit significant hardness-intensity and hardness-fluence 
correlations which match closely the correlations observed for GRBs.
They attribute this hardness evolution to an evolving $E_{peak}$.
The hardness ratio H $\equiv$\ S(50--100 keV)/S(25--50 keV) is close to unity for the
prompt emission and the first two flares, so all three of these
episodes fall well within the distribution of long bursts in a GRB
hardness-duration diagram \citep[see][for a recent BAT
hardness-duration distribution plot]{romi06}.
If the overall spectral shape of the flares remains roughly the same, 
we would expect the hardness ratio H to fall
dramatically as  $E_{peak}$\  shifts downward through the BAT energy
range (top panel, Figure~\ref{fig6}), since the BAT spectral index would shift from
the super-$E_{peak}$\ value of $\sim 1$\ to the sub-$E_{peak}$\ value of $\sim 2.5$.  This is indeed what we see in the
bottom panel of Figure~\ref{fig6}, supporting a common
origin for the flares.

Another interesting effect is seen in Figure~\ref{fig7}.  Here we see
that the prompt emission and the first two flares show a
relationship between $E_{\rm peak}$ and $E_{\rm\gamma,iso}$\ quite
consistent with the relationships found by \citet{amat02}.% and \citet{ghir04}. 
This supports a similar mechanism for these flares and
for the prompt emission,
since it suggests that these flares are spectrally and energetically
like weak, low $E_{\rm peak}$ bursts. The last three flares fall well
below the relations since for the succession of flares $E_{\rm peak}$
is falling more quickly than $E_{\rm\gamma,iso}$, with these flares
remaining relatively energetic ($E_{\rm\gamma,iso} > 10^{51}\;$erg,
even as $E_{\rm peak}$ decreases to $\lesssim 1\;$keV). This argues
against a similar origin for these two flares and the prompt emission
(especially since almost all known outliers to the Amati relation are
on the other side of it, i.e. with a higher $E_{\rm peak}$ for their
$E_{\rm\gamma,iso}$ than expected by this relation, while here the
situation is the opposite), although there are still possible ways
around this.  
Alternatively, one could certainly argue that such a wide dispersion within a 
single burst over a wide range of energies should call for a re-evaluation of 
the Amati relationship. Although the $E_{\rm peak} -
E_{\rm\gamma,iso}$\ relations have been shown to hold for many X-ray
flashes \citep[e.g.,][]{saka06a}, it is much less clear whether the
relations should hold for late-time flares or individual peaks within bursts.

The time structure of the flares is quite similar to that
of the prompt emission.  All of the flares show structure on time
scales of $\sim 10^{-1}\ -\ 10$\ s within an overall envelope of
duration $\sim 10\ -\ 50$\ s.  The prompt emission falls within a
$\sim 30$\ s envelope, and although the sub-peaks in the prompt
emission are not formally significant, there is a suggestion of
structure on shorter time scales, especially in the 50-100 keV band
(middle panel of Figure~\ref{fig1}).  Certainly short time-scale
variations in the prompt emission are a common feature of GRBs.

\section{Conclusions}

\noindent The temporal properties of the series of five flares in the
long gamma-ray burst GRB~060714 provide strong evidence against an
external shock (or a long lived reverse shock) origin for these X-ray
flares, and are consistent with sporadic late time activity of the
central source.
The strongest argument against an external shock origin of the flares
is that large amplitude variations in flux ($\Delta F \gtrsim F$)
occur on very short time scales ($\Delta t / t \ll 1$, where initially
$\Delta t / t \lesssim 10^{-2}$, while for all flares or sub-flares
$\Delta t / t < 0.1$; see Figs.~\ref{dtt_dff} and \ref{fig8}). As
discussed in \S\ref{extshock}, this cannot be reasonably accounted for
by emission from the forward shock, and similar considerations apply
also for a long lived reverse shock. 

In the context of the internal shocks model, we show that the temporal
properties of the flares ($\Delta t / t \ll 1$) exclude scenarios
where the colliding shells are ejected from the source well before the
observed times of the flares (see \S~\ref{ISa}), and require instead
intermittent late time activity of the central source. Even in the
latter picture, an internal shocks model with small contrast in the
Lorentz factor is inconsistent with the observations
 (see \S~\ref{ISa}), for the following
reason. The large $E_{\rm\gamma,iso}$\ of the five flares, together
with the small efficiency of such a model in converting the bulk
kinetic energy into the observed radiation, requires that a large
amount of kinetic energy remain in the shells, a condition that would
unavoidably produce very prominent ``refreshed shocks.''  Such
episodic energy injections into the afterglow shock are, however,
inconsistent with the temporal properties of the smooth X-ray (and
optical) light curve that follows the flares.

An internal shocks model involving sporadic late-time activity of the
central source and a reasonably large contrast in the Lorentz factors
of the colliding shells (\S~\ref{ISb}) is more consistent with the
data.  This type of model can avoid prominent ``refreshed shocks,'' in
addition to being able to account for both the observed temporal
properties of the flares and the decrease with time of their $E_{\rm
peak}$\ and $E_{\rm\gamma,iso}$. Although the data from GRB~060714 are
not sufficient for us to conclusively discriminate between particular
models of late-time intermittent activity of the central source
(\S\ref{relation}), they can start testing and perhaps even eventually
discriminate between the different models. For example, the viscous
disk model \citep{per06} can in principal explain the trend of peaks
becoming less sharp with increasing time.  The simplest expectations
of fragmentation models \citep{king05,burro05b}, however, are
inconsistent with the decrease in the Lorentz factor $\gamma$\ that
follows from the time decay of $E_{\rm peak}$\ and $E_{\rm\gamma,iso}$
in the internal shocks model. Moreover, the decline in
$E_{\rm\gamma,iso}$\ is steeper than what would be expected from
fallback in the collapsar model \citep{MWH01}, but suggestive of a
relativistic outflow powered by neutrino-anti neutrino annihilation.

Some evidence in favor of a common origin for the flares and the
prompt emission is provided by the spectral and temporal similarities
of these components. As in the prompt emission, there is a clear
spectral softening as the flares progress, as well as within most of
the flares. Furthermore, both the prompt emission and the flares show
time structure on the scale of $\sim 1$\ s within an overall envelope
of duration $\sim 10$\ s. On the other hand, an arguments against a
common origin is the inconsistency of the final three flares with the
\citet{amat02} %and \citet{ghir04} 
relation, although it is not well-established that this relation should
hold for late-time flares.

After the fifth and final flare, GRB~060714 showed typical afterglow
time profiles consisting of three separate power law segments
(Figure~\ref{fig2}).  This part of the light curve is a nice
example of the ``canonical'' afterglow light curve found by {\it Swift}
\citep{nou06,panait06,zhang06}. This burst was well observed at late
times in the optical (Figure~\ref{fig3}) and the optical light curve
shows both similarities and differences with respect to the X-ray
light curve. The first optical observations showed emission at a
typical value for an optical afterglow in the {\it Swift} era (V =
18.6 at $\sim 200$\ s), and the optical magnitude remained roughly
constant out to at least $T_0+ 7700$\ s, with no indication of an
optical counterpart to any of the flares. Similar to the X-ray light
curve, the R-band light curve also had a break to a steeper temporal
decay, although the optical break
occurred about a factor of $\sim 2$ in time later than the X-ray break.
Both before and after the break, the X-ray and R-band power law
temporal decay indices were quite similar. This suggests a similar
origin for the late-time X-ray and optical emission --- most likely
afterglow emission from the forward shock, probably with some
contribution from a long lived reverse shock before and around the
time of the break, which might potentially account for the difference
in the breaks times between the optical and the X~rays.

The subject of X-ray flares in GRBs is an important one and a number of recent papers on {\it Swift} bursts have discussed flares, leading to a growing consensus that flares reflect late time activity of the central source.  Some important examples are \citet{falc06} (the giant flare of GRB~050502B),  \citet{guet07} (GRB~050713A), and \citet{burro07} and \citet{chin07} on the analysis of a  statistical sample of {\it Swift} flares.  This paper adds important new observational results and conclusions to the existing body of literature on flares. GRB~060714 contains an unusually large number of well monitored flares in a single event, and our interpretation of the timing and energetics of these flares not only provides strong evidence against an external shock or a long lived reverse shock model, but allows us to favor the model of late-time activity of the central source and a large contrast in the Lorentz factor of the colliding shells over an internal shocks model with a small contrast in the Lorentz factors and refreshed shocks.  This event provides the best evidence to date of a continuous and
gradual transition in the spectral and temporal properties of the prompt
emission and flares, which is very suggestive of a common origin.

\acknowledgments We gratefully acknowledge the many useful comments and 
suggestions from G. Chincarini and useful discussions with T. Sakamoto.
HAK was supported in this work by the Swift project,
funded by NASA. DNB and DM are supported by NASA contract NAS5-00136.
JG was supported by the U.S. Department of
Energy under contract DE-AC03-76SF00515.

\newpage
\appendix
\section{Derivation of the Equations in \S~\ref{ISa}}
\label{low_contrast_IS}

Here we provide a derivation of the equations for internal shocks with
a small contrast in the Lorentz factor, that appear in
\S~\ref{ISa}. Consider two shells, where the (back edge of the) first
shell (subscript `1') is ejected from the source at (lab-frame) time
$t_1 = t_{\rm ej}$ with Lorentz factor $\gamma_1 = \gamma$ and rest
mass $m_1$, while the (front edge of the) second shell (subscript `2')
is ejected at $t_2 = t_{\rm ej} + \Delta t_{\rm ej}$ with a Lorentz
factor $\gamma_2 = \gamma+\Delta\gamma$ and rest mass $m_2$. We are
interested in the limit of low contrast in the Lorentz factors of the
two shells, $\Delta\gamma/\gamma \ll 1$, and provide below approximate
expressions which are valid in that limit. The two shells collide at a
lab frame time $t_{\rm IS}$ and a radius $R_{\rm IS}$ which satisfy
$R_{\rm IS} = R_1(t_{\rm IS}) = R_2(t_{\rm IS})$ where $R_1(t) =
\beta_1c(t-t_{\rm ej})$ and $R_2(t) = \beta_2c(t-t_{\rm ej} - \Delta
t_{\rm ej})$ so that $t_{\rm IS} = t_{\rm ej} + \beta_2\Delta t_{\rm
ej}/(\beta_2-\beta_1)$ and
\begin{equation}\label{R_IS}
R_{\rm IS} = \frac{\beta_1\beta_2c\Delta t_{\rm ej}}{\beta_2-\beta_1}
\cong \frac{2\gamma_1^2c\Delta t_{\rm ej}}{1-(\gamma_1/\gamma_2)^2}
\approx \frac{\gamma}{\Delta\gamma}\,\gamma^2c\Delta t_{\rm ej}\ ,
\end{equation}
where the ``$\cong$'' is valid for $\gamma_1,\,\gamma_2\gg 1$, while
the ``$\approx$'' requires also $\Delta\gamma/\gamma \ll 1$.

The observed time corresponding to the onset of the resulting spike in
the light curve (or ``flare'') is the arrival time of a photon emitted
along the line of sight at $R_{\rm IS}$ and $t_{\rm IS}$, relative to
a photon emitted at $R = 0$ and $t = 0$, which is given by
\begin{equation}
t_{\rm flare,onset} = t_{\rm IS} - \frac{R_{\rm IS}}{c} = 
t_{\rm ej} + \frac{\beta_2(1-\beta_1)\Delta t_{\rm ej}}{\beta_2-\beta_1}
\cong t_{\rm ej} + \frac{\Delta t_{\rm ej}}{1-(\gamma_1/\gamma_2)^2}
\approx t_{\rm ej}+\frac{\gamma}{\Delta\gamma}\,\frac{\Delta t_{\rm ej}}{2}\ .
\end{equation}
The exact observed time at which the flare peaks, $t_{\rm flare}$,
would be somewhat later, typically by about the shell shock crossing
time. The latter depends on the details of the shell (width, Lorentz
factor and mass density distribution within the shell, etc.) but is
generally expected to be of the order of $t_{\rm flare,onset} - t_{\rm
ej}$. Furthermore, the typical width of the spike in the light curve,
$\Delta t_{\rm flare}$, is of the order of the angular time (which is
typically also of the order of the radial time or shock crossing
time), and since $\Delta R_{\rm IS} \sim R_{\rm IS}$ we have
\begin{equation}
\Delta t_{\rm flare} \sim \frac{R_{\rm IS}}{c\gamma^2} 
\approx \frac{\gamma}{\Delta\gamma}\Delta t_{\rm ej} 
\sim t_{\rm flare} - t_{\rm ej} \sim 
2(t_{\rm flare,onset} - t_{\rm ej})\ .
\end{equation}
In order for $\Delta\gamma$ to be meaningful, we assume that it is
larger than the spread in the Lorentz factor within each shell, so
that the shells do not spread significantly before $R_{\rm IS}$ and
their width (in the lab frame) is of order $c\Delta t_{\rm ej}$.

The values of the spectral break frequencies depend on the physical
conditions within the shocked shells. The isotropic equivalent kinetic
luminosity of the outflow is given by $L_{\rm iso} \approx 4\pi R^2
\gamma^2\rho' c^3$ while the total luminosity is $(1+\sigma)L_{\rm
iso}$ where $\sigma = (B')^2/4\pi\rho'c^2$ is the ratio of
electromagnetic to kinetic energies of the outflow, and primed
quantities are measured in the comoving frame. The relative velocity
between the two shells is
\begin{equation}
\beta_{21} = \frac{\beta_2-\beta_1}{1-\beta_2\beta_1} \cong
\frac{\gamma_2^2-\gamma_1^2}{\gamma_2^2+\gamma_1^2} \approx 
\frac{\Delta\gamma}{\gamma} \ll 1\ .
\end{equation}
Thus, the shocks going into the two colliding shells are Newtonian for
$\Delta\gamma/\gamma \ll 1$, and the compression ratio is $\rho'_{\rm
ps}/\rho' \approx 4$, where the subscript ``ps'' is for
post-shock. For equal density shells, the relative velocity of the
upstream and downstream fluids is $\beta_{\rm ud} \approx \beta_{21}/2
\approx \Delta\gamma/2\gamma$, and the internal energy per unit
rest-energy is~\footnote{In this case, if the mass of the shells is
also the same, the two shocks finish crossing the two shells together,
and this is also the fraction $\epsilon$ of the total energy that is
converted into internal energy. As shown below, for a fixed Lorentz
factor contrast $\Delta\gamma/\gamma$, $\epsilon$ is maximal for equal
mass shells.}  $e'_{\rm ps}/\rho'_{\rm ps}c^2 \approx \beta_{\rm
ud}^2/2 \approx (\Delta\gamma/\gamma)^2/8$. Thus, the minimal random
Lorentz factor of the power-law distribution of relativistic
electrons scales as
\begin{equation}
\gamma_m \propto \epsilon_e\frac{e'_{\rm ps}}{\rho'_{\rm ps}} \propto 
\epsilon_e\left(\frac{\Delta\gamma}{\gamma}\right)^2\ ,
\end{equation}
where $\epsilon_e$ is the fraction of the post-shock internal energy
that goes into such a population of relativistic electrons.  The
internal energy in the shocked regions scales as
\begin{equation}
e'_{\rm ps} \sim \frac{1}{8}\left(\frac{\Delta\gamma}{\gamma}\right)^2\rho'c^2 
\propto \left(\frac{\Delta\gamma}{\gamma}\right)^2
\frac{L_{\rm iso}}{\gamma^2R^2} 
\propto \left(\frac{\Delta\gamma}{\gamma}\right)^4 
\frac{L_{\rm iso}}{\gamma^6(\Delta t_{\rm ej})^2}\ ,
\end{equation}
where we evaluate the values of the relevant quantities at $R \approx
R_{\rm IS}$, using Eq.~\ref{R_IS}.

If the magnetic field holds a constant fraction, $\epsilon_B$, of the
internal energy behind the shock, as is often assumed for a
shock-generated magnetic field, then such an equipartition
field~\footnote{We use the term ``equipartition field'' for
simplicity, even though strictly speaking it is holds a constant
fraction ($\epsilon_B^{1/2}$, generally smaller than unity) of the
equipartition value.}  would scale as
\begin{equation}
B'_{\rm eq} \propto \left(\epsilon_Be'_{\rm ps}\right)^{1/2} \propto 
\epsilon_B^{1/2}\left(\frac{\Delta\gamma}{\gamma}\right)^2
L_{\rm iso}^{1/2}\gamma^{-3}(\Delta t_{\rm ej})^{-1}\ .
\end{equation}
However, some magnetic field is expected to be advected with the
outflow from the central source. At large distances form the source
such a field is expected to be primarily in the tangential direction
(normal to the radial direction) so that it would be amplified in the
shock by the compression ratio, $B'_{\rm adv,ps} \approx 4B'_{\rm
adv}$ where $(B'_{\rm adv})^2 = 4\pi\rho'c^2\sigma$. Thus, for equal
mass shells the ratio of the magnetic energy density associated with
this field to the internal energy density in the shocked region is
$(B'_{\rm adv,ps})^2/8\pi e'_{\rm ps} \approx
64\sigma(\Delta\gamma/\gamma)^{-2}$, so that the shock compressed
advected field would typically exceed an equipartition shock generated
field for reasonable values of $\sigma$, i.e. $\sigma \gtrsim
10^{-3}(\Delta\gamma/\gamma)^2(\epsilon_B/0.1)$. In this case $B'_{\rm ps}
\approx B'_{\rm adv,ps}$, where
\begin{equation}
B'_{\rm adv,ps} \approx 8c\sqrt{\pi\rho'\sigma}
\propto \sigma^{1/2}\left(\frac{\Delta\gamma}{\gamma}\right)
L_{\rm iso}^{1/2}\gamma^{-3}(\Delta t_{\rm ej})^{-1}\ .
\end{equation}
The synchrotron frequency of the electrons with the minimal random
Lorentz factor, $\gamma_m$, scales as 
\begin{equation}\label{nu_m}
\nu_m \approx \gamma\frac{eB'_{\rm ps}\gamma_m^2}{2\pi m_e c}
\propto \left\{\matrix{\epsilon_B^{1/2}
\left(\frac{\Delta\gamma}{\gamma}\right)^6\epsilon_e^2 L_{\rm iso}^{1/2}
\gamma^{-2}(\Delta t_{\rm ej})^{-1} & \quad ({\rm equipartition})\ ,\cr \cr
\sigma^{1/2}
\left(\frac{\Delta\gamma}{\gamma}\right)^5\epsilon_e^2 L_{\rm iso}^{1/2}
\gamma^{-2}(\Delta t_{\rm ej})^{-1} & \quad ({\rm advected\ field})\ .} \right.
\end{equation}
The random Lorentz factor of electrons that cool on the dynamical time
(shell crossing time), $t' \sim R_{\rm IS}/\gamma$, scales as
\begin{equation}
\gamma_c \approx 
\frac{6\pi m_e c}{\sigma_T (B'_{\rm ps})^2(R_{\rm IS}/\gamma)(1+Y)}
\propto \left\{\matrix{(1+Y)^{-1}\epsilon_B^{-1}
\left(\frac{\Delta\gamma}{\gamma}\right)^{-3}L_{\rm iso}^{-1}
\gamma^5\Delta t_{\rm ej} & \quad ({\rm equipartition})\ ,\cr \cr
(1+Y)^{-1}\sigma^{-1}\left(\frac{\Delta\gamma}{\gamma}\right)^{-1}
L_{\rm iso}^{-1}\gamma^5\Delta t_{\rm ej} & 
\quad ({\rm advected\ field})\ ,} \right.
\end{equation}
where $Y$ is the Compton y-parameter. 

The factor of $(1+Y)$ can be safely dropped from the expression for
the advected field, since in this case $Y \ll 1$ (it is included here
for completeness). This can be seen as follows. In general $Y(1+Y)
\approx \beta_{2,{\rm sh}}\epsilon_{\rm rad}\epsilon_e/\epsilon_B$
\citep[see, e.g.][]{SE01} where $\beta_{2,{\rm sh}}$ is the velocity
of the downstream medium relative to the shock front, which in our
case is $\beta_{2,{\rm sh}} \sim \beta_{21} \approx
\Delta\gamma/\gamma \ll 1$, and $\epsilon_{\rm rad} \sim
\min[1,(\nu_m/\nu_c)^{(p-2)/2}]$ is the fraction of the energy in the
post-shock power law distribution of relativistic electrons that is
radiated away. Here $\epsilon_B = (B'_{\rm ps})^2/8\pi e'_{\rm ps}$
which for a sub-equipartition field is $< 1$, but for a field advected
from the source and compressed by the shock it is (see above) $\approx
64\sigma(\Delta\gamma/\gamma)^{-2}$ which is $\gg 1$ unless the
magnetization of the outflow is extremely low, $\sigma \lesssim
10^{-2}(\Delta\gamma/\gamma)^2$. Thus $Y \sim \epsilon_e\epsilon_{\rm
rad}10^{-2}\sigma^{-1}(\Delta\gamma/\gamma)^3 \ll 1$. For a
shock-generated field that is a constant fraction of equipartition $Y$
would also be small unless $\epsilon_{\rm rad}\epsilon_e/\epsilon_B
\gtrsim \gamma/\Delta\gamma$.

The cooling break frequency thus scales as
\begin{equation}\label{nu_c}
\nu_c \approx \gamma\frac{eB'_{\rm ps}\gamma_c^2}{2\pi m_e c}
\propto \left\{\matrix{(1+Y)^{-2}\epsilon_B^{-3/2}
\left(\frac{\Delta\gamma}{\gamma}\right)^{-4}L_{\rm iso}^{-3/2}
\gamma^8\Delta t_{\rm ej} & \quad ({\rm equipartition})\ ,\cr \cr
(1+Y)^{-2}\sigma^{-3/2}\left(\frac{\Delta\gamma}{\gamma}\right)^{-1}
L_{\rm iso}^{-3/2}\gamma^8\Delta t_{\rm ej}
 & \quad ({\rm advected\ field})\ .} \right.
\end{equation}

Finally, we derive the fraction, $\epsilon$, of the total energy that
is converted into internal energy during the collision between the two
shells (in the limit $\sigma \ll 1$ and $\Delta\gamma/\gamma \ll
1$). Conservation of energy and momentum read
\begin{eqnarray}
\gamma_1m_1 + \gamma_2m_2 &=& \gamma_f M\ ,
\\
\gamma_1\beta_1m_1 + \gamma_2\beta_2m_2 &=& \gamma_f\beta_f M\ ,
\end{eqnarray}
where $\gamma_f = (1-\beta_f^2)^{-1/2}$ is the final Lorentz factor,
and $M = m_1 + m_2 + E_{\rm int}'/c^2$ where $E'_{\rm int}$ is the
internal energy that was produced in the collision, as measured in the
rest frame of the merged shell, while its value in the lab frame is
$E_{\rm int} = \gamma_f E'_{\rm int}$. One obtains $\beta_f$ from the
ratio of the two equations, and for $\gamma \gg 1$ we have
\begin{equation}
\frac{1}{2\gamma_f^2} \cong 1-\beta_f = 
\frac{\gamma_1(1-\beta_1)m_1 + \gamma_2(1-\beta_2)m_2}
{\gamma_1m_1 + \gamma_2m_2} \cong
\frac{m_1/\gamma_1 + m_2/\gamma_2}{2(\gamma_1m_1+\gamma_2m_2)}\ ,
\end{equation}
and therefore
\begin{equation}
\gamma_f \cong
\sqrt{\frac{\gamma_1m_1+\gamma_2m_2}{m_1/\gamma_1+m_2/\gamma_2}}\ .
\end{equation}
Thus we have
\begin{equation}
\epsilon = \frac{E_{\rm int}}{(\gamma_1m_1+\gamma_2m_2)c^2} =
1-\frac{\gamma_f(m_1+m_2)}{\gamma_1m_1+\gamma_2m_2} \cong
1-\left[1+\frac{m_1m_2}{(m_1+m_2)^2}\left(\frac{\gamma_2}{\gamma_1}
+\frac{\gamma_1}{\gamma_2}-2\right)\right]^{-1/2}\ ,
\end{equation}
and since for $\Delta\gamma/\gamma \ll 1$ (in addition to $\gamma \gg
1$),
\begin{equation}
2(\gamma_{21}-1) \cong \frac{\gamma_2}{\gamma_1}
+\frac{\gamma_1}{\gamma_2}-2 \approx
\left(\frac{\Delta\gamma}{\gamma}\right)^2 \ll 1\ ,
\end{equation}
then
\begin{equation}
\epsilon \approx \frac{m_1m_2}{2(m_1+m_2)^2}
\left(\frac{\Delta\gamma}{\gamma}\right)^2
= \frac{x}{2(1+x)^2}\left(\frac{\Delta\gamma}{\gamma}\right)^2\ ,
\end{equation}
where $x = m_2/m_1$ (or alternatively $m_1/m_2$) is the rest-mass
ratio of the two shells.

%% Tables should be submitted one per page, so put a \clearpage before

%% Two options are available to the author for producing tables:  the
%% deluxetable environment provided by the AASTeX package or the LaTeX
%% table environment.  Use of deluxetable is preferred.
%%

%% Three table samples follow, two marked up in the deluxetable environment,
%% one marked up as a LaTeX table.

%% In this first example, note that the \tabletypesize{}
%% command has been used to reduce the font size of the table.
%% We also use the \rotate command to rotate the table to
%% landscape orientation since it is very wide even at the
%% Note also that the \label command needs to be placed

%% This table also includes a table comment indicating that the full
%% version will be available in machine-readable format in the electronic

\clearpage

\begin{deluxetable}{cccccc}
\tabletypesize{\scriptsize}
\tablecaption{Data displayed in the top three panels of Figure~\ref{fig6}\label{tbl-3}}
\tablewidth{0pt}
\tablehead{
\colhead{Time Interval} & \colhead{Fluence (0.3-10 keV)} & \colhead{Fluence (15-150 keV)} & \colhead{$E_{peak}$\ (keV)} & \colhead{Power Law index} & \colhead{$E_{iso}\ ({\rm 10^{52} ergs})$}}
\startdata
-13.4 -- 18.0 &  -- &  $3.95^{+0.45}_{-0.89}$  & $ > 77.6 $ & 1.61 $\pm$ 0.13 & $ > 3.74$ \\
      70.21   --  86.2 &  --  &   $3.44^{+0.35}_{-2.45}$  & 55.5  $\pm$    11.5 & 1.00 $\pm$ 0.51 & $1.36^{+0.55}_{-0.29}$\\
      86.2    --  102.88 & --   & $1.90^{+1.68}_{-0.49}$  &   45.0  $\pm$    9.6 & 1.38 $\pm$ 0.36 & $1.76^{+0.74}_{-0.39}$ \\
      107.0   --   121.04 & 1.35$\pm$ 0.16  & 0.66 $\pm$ 0.15 & 9.8  $\pm$    1.5 &  1.50 $\pm$ 0.16 & 0.64 $\pm$ 0.07\\
      121.04   --   159.21 & $0.58^{+0.06}_{-0.07}$   &  $0.06^{+0.06}_{-0.04}$   & 3.9  $\pm$    0.77 & 1.51 $\pm$ 0.24 & 0.58 $\pm$ 0.10\\
      159.21   --   199.21 & $0.32^{+0.34}_{-0.09}$   & $< 0.05
      04$ &  0.5   $\pm$    0.50 & 1.90  $\pm$ 0.26 & 0.36 $\pm$0.07 \\
\enddata
%% Text for table notes should follow after the \enddata but before
%% the \end{deluxetable}. Make sure there is at least one \tablenotemark
%% in the table for each \tablenotetext.
\tablecomments{Fluence is in units $10^{-8} erg\ cm^{-2}$}
%tablenotetext{b}{Another sample footnote for table~\ref{tbl-1}}
\end{deluxetable}

\begin{deluxetable}{cccccccc}
\tabletypesize{\scriptsize}
\tablecaption{Temporal decay slopes of flares\label{tbl-1}}
\tablewidth{0pt}
\tablehead{
\colhead{Flare} & \colhead{Peak} & \colhead{Start $t_1$ } &
\colhead{Rise time $\Delta t$ (s)} & \colhead{Decay index $\alpha_{\rm A}$} 
& \colhead{$t_0$ for $\alpha_{\rm B}$\ (s)} & 
\colhead{Decay index $\alpha_{\rm B}$}}
\startdata
\multicolumn{2}{l}{prompt} & -13.4 & 15.5 & 5.4 $\pm$ 1.1 & -13.17 & 2.34 $\pm$ 0.45\\
\multicolumn{2}{l}{prompt decay} & 18.0 & -- & 2.1 $\pm$ 0.8 & -13.17 & 1.16 $\pm$ 0.47\\
1 & 1 & 70.2 & 0.19 & 16.3 $\pm$ 5.3 & 70.21 & 0.29 $\pm$ 0.11\\
1 & 2 & 73.6 & 2.01 & 24.0 $\pm$ 4.1 & 70.21 & 2.14 $\pm$ 0.36\\
1 & 3 & 78.8 & 0.39 & 23.9 $\pm$ 2.8 & 70.21 & 3.35 $\pm$ 0.39\\
2 & 1 & 86.2 & 1.77 & 19.2 $\pm$ 2.5 & 86.2  & 1.33 $\pm$ 0.18\\
2 & 2 & 95.5 & 3.54 & 23.7 $\pm$ 4.8 & 86.2  & 3.50 $\pm$ 0.71\\
3 & 1 & 102.9 & 2.54 & 46.9 $\pm$ 17.7 & 108.5 & 3.77 $\pm$ 1.42\\
3 & 2 & 118.3 & 3.14 & 41.2 $\pm$ 18.2 & 108.5 & 4.56 $\pm$ 2.01\\
4 & 1 & 122.5 & 11.2 & 10.3 $\pm$ 5.1 & 122.54 & 0.96 $\pm$ 0.46\\
4 & 2 & 136.2 & 2.52 & 10.7 $\pm$ 0.5 & 122.54 & 1.78 $\pm$ 0.08\\
5 & -- & 160.7 & 18.0 & 17.7 $\pm$ 0.7 & 160.71 & 2.73 $\pm$ 0.12\\
\multicolumn{2}{l}{final prompt decay} & 195.6 & --- & 6.34 $\pm$ 0.39 & 160.71 &  2.14 $\pm$ 0.13\\
\multicolumn{2}{l}{shallow afterglow} & 323.8 & --- & 0.31 $\pm$ 0.17 & 160.71 &  0.24 $\pm$ 0.03\\
\multicolumn{2}{l}{steep afterglow} & 3200 & --- & 1.24 $\pm$ 0.05 & 160.71 & 1.21 $\pm$ 0.03\\
\enddata
%% Text for table notes should follow after the \enddata but before
%% the \end{deluxetable}. Make sure there is at least one \tablenotemark
%% in the table for each \tablenotetext.
\tablecomments{Flares are numbered as in the text and Figures~\ref{fig1} and \ref{fig4}.  Times are with
reference to the trigger time $T_0$\ and the definition of the rise
time is given in the text.  The decay index is derived through a
fit to: $R \propto (t-t_0)^{-\alpha}$, where $R$ is the photon event
rate, $t$ is the time, and $t_0$\ is defined as the trigger time
$T_0$\ when deriving $\alpha_{\rm A}$\ (column 5), and as the start of the
particular peak or flare when deriving $\alpha_{\rm B}$\ (column 7).  The
values of $t_0$\ used in the derivation of $\alpha_{\rm B}$\ are shown in
column 6.}
%\tablenotetext{a}{Sample footnote for table~\ref{tbl-1} that was generated
%with the deluxetable environment}
%tablenotetext{b}{Another sample footnote for table~\ref{tbl-1}}
\end{deluxetable}

\newpage

\newpage

\newpage

\newpage

\newpage
\clearpage
%\plotone{f6.eps}

\newpage

\newpage

\newpage

\newpage

\end{document}